\documentclass{GleghornLab-StyleBioRxiv}
\usepackage{blindtext}
\usepackage{url}
\usepackage[many]{tcolorbox}
\usepackage{listings}
\usepackage{lscape}
\usepackage{wrapfig}
\usepackage{longtable}
\usepackage{subcaption}
\usepackage{float}
\usepackage{calc}
\usepackage{multirow}
\usepackage{booktabs}
\usepackage{tabularx}
\usepackage{seqsplit}
\usepackage{ragged2e}
\usepackage{array}
\usepackage{nicefrac}
\usepackage{microtype}
\usepackage{rotating}
\usepackage{colortbl}

\leadauthor{Hallee}

\begin{document}

\newtcolorbox{boxA}{
   %fontupper = \color{blue}, % font color
  % \colorlet{LightLavender}{Lavender!40!}
    boxrule = 1.5pt,
   % colframe = main,
    rounded corners,
    colback=blue!10!white
    %arc = 5pt   % corners roundness
}

\newtcolorbox{boxB}{
 %  fontupper = \bf\color{grey}, % font color
    boxrule = 1.5pt,
   % colframe = main,
    rounded corners,
    arc = 5pt   % corners roundness
}

\title{Diffusion Sequence Models for Enhanced Protein Representation and Generation}
\shorttitle{Diffusion Sequence Models for Enhanced Protein Representation and Generation}

\author[1, 2]{Logan Hallee}
\author[1]{Nikolaos Rafailidis}
\author[2]{David B. Bichara}
\author[1, 2, 3 \Letter]{Jason P. Gleghorn}
\affil[1]{{Center for Bioinformatics and Computational Biology, University of Delaware}}
\affil[2]{{Synthyra, PBLLC}}
\affil[3]{{Department of Biomedical Engineering, University of Delaware}}
\maketitle

\begin{abstract}
Proteins are fundamental to biology, executing diverse functions through complex physicochemical interactions, and they hold transformative potential across medicine, materials science, and environmental applications. Protein Language Models (pLMs) aim to unlock insights from the vast space of unlabeled protein sequences by learning rich, semantic representations from primary sequences via masked language modeling. However, these models typically exhibit limited generative capacity. In this work, we introduce the Diffusion Sequence Model (DSM), a novel pLM trained with masked diffusion to enable both high-quality representation learning and generative protein design. DSM builds upon the ESM2 architecture by incorporating a masked forward diffusion process inspired by the LLaDA framework. After training, DSM is capable of generating diverse, biomimetic sequences that align with expected amino acid compositions, secondary structures, and predicted functions, even with 90\% token corruption. Furthermore, DSM's learned representations match or exceed those of similarly sized pLMs on downstream tasks. We also introduce DSM$_{ppi}$, a variant fine-tuned to generate protein binders by attending to target sequences. We demonstrate DSM$_{ppi}$'s effectiveness on the challenging Bench-tested Binder Benchmark (BenchBB), where both DSM and DSM$_{ppi}$ produce candidates with superior predicted binding affinity compared to known binders. Our results establish masked diffusion as a powerful paradigm for unifying protein representation and generation in a single framework.

\end{abstract}

\vspace{0.25em}
\begin{corrauthor}
gleghorn\at udel.edu
\end{corrauthor}

\section{Introduction}
The evolution of protein systems occurs through gradual, mostly deleterious, and seemingly stochastic modifications to genetic sequences. The niche adaptation of biological systems via selection narrows the natural protein landscapes into a minuscule portion of the possible protein universe \cite{Shen_song_li_Zhang_2022, Loewe_Hill_2010, AV}. In fact, exploring the total protein universe appears infeasible by any natural process with over $10^{1301}$ possible protein sequences of length 2,048. Even within the smaller space of known natural proteins, we understand a small portion mechanistically. A fraction of 1\% of cataloged protein sequences have ever been expressed, and fewer have been well annotated \cite{uniprot, jones_estimating_2007, silveira_enzymap_2014, AV}. This disparity highlights the scarcity of experimental data, rendering protein design and functional annotation persistent challenges in the life sciences.

A deeper understanding of the protein universe is critical for advancing knowledge of disease mechanisms and biological systems \cite{markus_accelerating_2023, gleghorn_synthetic_2019, gleghorn_3d_2022, gleghorn_virus_2022, gleghorn_maternal_2021, Gilbert_Gleghorn_2023, Shen_Gleghorn_2025, roberts_potential_2021, sen_potential_2021}. Moreover, designed proteins have other transformative impacts beyond biology, with potential to advance plastic degradation and recycling, carbon capture, and the creation of novel materials \cite{herrero_acero_surface_2013, pang_catalytic_2020, miserez_protein-based_2023}. Unfortunately, the experimental validation of protein properties remains time- and money-intensive; the same can be said about traditional protein design. Therefore, there is a crucial need for reliable computational systems that can annotate and generate protein sequences. % maybe cite about costs

In response, the Protein Language Model (pLM) community has pursued these aims in tandem with the rapid advancement of natural language processing (NLP) systems. By treating proteins as a semantic language, various transformer-like neural networks are an effective choice to map gene products to meaningful numerical representations. Through amino acid, codon, nucleotide, or atomic inputs, researchers typically deploy semi-supervised masked language modeling (MLM) to pretrain models \cite{elnaggar_prottrans_2022, hallee_cdsbert_2023, li_codonbert_2023, ren_codonbert_2024, nguyen_sequence_2024, zheng_esm_nodate, evo2}. The linear primary sequences of proteins fold into intricate 3D structures, so the global interaction relationships mapped with bidirectional attention have dominated as a modeling choice. The combination of bidirectional attention and MLM naturally encodes the protein structural information into the attention maps of pLMs when a large enough layer depth is used, offering an emergent phenomenon that enhances the representations from pLMs \cite{esm1}. However, while pLM representations can be used and fine-tuned for many downstream tasks, they do not inherently map well to abstract concepts like biological processes or fitness \cite{li_feature_2024, AV}. Additionally, they have poor generative modeling capabilities because they are typically only trained to denoise inputs from 15-20\% noise \cite{elnaggar_ankh_2023, AV}. While autoregressive (AR) models offer a solution to generative capabilities, AR-based pLMs typically have worse representations on downstream tasks \cite{dplm, is_scaling}. Additionally, we argue that it is an ineffective modality for protein generation. While the linear left-to-right generative process mimics protein synthesis well, the phenotypic changes from mutagenesis and epistasis occur based on the conditional probability of global context. As such, implementation of bidirectional token information mixing is a beneficial choice \cite{dplm, lyra, is_scaling}.

To address the limitations of generative pLMs, we propose Diffusion Sequence Modeling (DSM), a framework that unifies biologically meaningful representation and generative modeling through masked diffusion. To achieve this, we modified the Large Language Diffusion Models (LLaDA) framework \cite{llada} to further pretrain popular pLMs using a masked-based diffusion process. DSM extends ESM2 \cite{lin_evolutionary-scale_2023} with a novel language modeling head and training objective, enabling robust denoising across high corruption rates and sequence generation with global context. After training, we observed low cross-entropy losses and accurate sequence reconstruction at low mask rates up to 90\% corruption. With unconditional generative sampling, DSM models produced distinct sequence distributions that closely mimicked amino acid k-mers, predicted secondary structures, and predicted protein functions compared to natural sequences. In addition to their generative prowess, DSM models also produced high-quality protein sequence representations. DSM models match or outperform MLM-based and discrete diffusion pLMs (DPLM) of the same size as well as an AR pLM almost twice DSM's size. We further extended the DSM framework by finetuning DSM on sets of interacting protein pairs with the hope of leveraging target protein inputs to generate protein binders (DSM$_{ppi}$). Our case study with DSM$_{ppi}$ and DSM highlighted the potential of unconditional and conditional generation schemes for protein design, with both methods producing promising protein binders with higher predicted binding affinity than the best known binders in BenchBB \cite{egfr_comp}. BenchBB provides a rigorous benchmark by featuring diverse protein targets representative of therapeutically relevant binder design, directly demonstrating DSM$_{ppi}$’s practical applicability \cite{egfr_comp}. Together, these results establish masked diffusion as a natural fit for pLMs, offering a unified framework for high-quality representation and biologically coherent generation.

\section{Background}

\subsection{LLaDa}
The LLaDA framework introduces a diffusion-based alternative to standard AR generative modeling in NLP by training a transformer neural network with a masked diffusion forward process \cite{llada}, offering bidirectional context utilization in contrast to AR models. This makes it particularly appealing for protein sequences, where functional properties often exhibit long-range dependencies in 1D but are actually close in 3D, only fully realized in forward passes with bidirectional context. More formally, LLaDa training assumes a neural network with parameters $\theta$, $p_\theta(\cdot,x_t)$, that takes $x_t$ as input and predicts all masked tokens $M$ in a single forward pass. The objective is
\[
\mathcal{L}(\theta)=
-\,\mathbb{E}_{t,x_{0},x_{t}}
\!\left[
\frac{1}{t + \epsilon}
\sum_{i=1}^{L}
\mathbf{1}\bigl[x_{t}^{i}=M\bigr]\,
\log p_{\theta}\!\bigl(x_{0}^{i}\mid x_{t}\bigr)
\right],
\]
where $x_0$ is sampled from the data with sequence length $L$ and $t \in [0, 1]$ is sampled uniformly, independently for each mini-batch. The indicator function $\mathbf{1}(\cdot)$ gathers only masked tokens, so that unmasked token logits are not included in the loss value \cite{llada}. Importantly, the $\frac{1}{t}$ steers away from the traditional MLM training objective by penalizing the model more severely for low mask examples, scaling the loss consistently based on the difficulty of the input. A small $\epsilon$ prevents division by 0. Notably, this generative format complies with \textit{Fisher consistency}, implying favorable scaling for larger models and datasets \cite{llada, fisher}.

After training, LLaDA models can perform a reverse process over a corpus of masked tokens. While the trained model unmasks all tokens every forward pass, some tokens can be remasked with various sampling methods to simulate a diffusion process. After fine-tuning, the eight billion parameter version of LLaDA performs on par with autoregressive large language models (LLMs) on several downstream benchmarks, including superior performance to LLaMA3 eight billion on MMLU, TruthfulQA, as well as math and coding tasks \cite{llada}.

\subsection{Annotation Vocabulary}

The Annotation Vocabulary (AV) is a standardized combination of various protein-related ontologies organized into a machine-readable format. The original version utilizes Enzyme Commission (EC) numbers, Gene Ontologies (GO) for Biological Process (BP), Molecular Function (MF), and Cellular Compartment (CC), as well as Interpro (IP) and Gene3D (3D) domains \cite{AV}. In this work, we expanded AV with UniProt keywords and cofactor information to further enhance the vocabulary \cite{uniprot}. AV tokens are mapped to unique integers, enabling direct token embedding for various representation learning schemas. AV has previously been used to train protein representation and generation models \cite{AV}. We annotated proteins in this work using the \textbf{Translator} model (\textbf{Supplemental \ref{sec:supp_Translator}}), and used AV terms for downstream evaluation \cite{AV}.

\subsection{Alignment Score}

Needleman-Wunsch global sequence alignment with evolutionarily informed scoring matrices is an ideal method for evaluating the similarity of two amino acid sequences, as it can recognize biologically similar motifs that are located at different positions. However, the score trends monotonically with the length of the inputs, providing an unscaled comparison that is not ideal for a standardized similarity metric. We previously developed a normalized Needleman-Wunsch score to design a more interpretable metric, called the Alignment Score (ASc) \cite{AV}. Defined as $ASc(a, b) = \frac{l}{f(a, a) - f(a, b) + l}$, where $f(a,b)$ is the Needleman-Wunsch alignment score with BLOSUM62 between ground truth sequence $a$ and generated sequence $b$, and $l$ is the length of sequence $a$. The result is an error term between intra- and inter-sequence alignment in the denominator that reduces the score upon poor alignment. The score ranges from 0 (no similarity) to 1 (same sequence), with random sequence pairs typically scoring around 0.15, and similarity gradually approaching 1 \cite{AV}. We use ASc throughout this work as a normalized measure of sequence recovery quality, especially at high corruption levels.

\section{Methods}
\label{sec:methods}

\subsection{Computational resources}

All experiments used a single A5000 GPU, A6000 GPU, GH200 GPU, or four A100 GPUs. The pretraining of DSM$_{650}$ was the most computationally intensive, requiring 12 days of 4xA100 time. The remaining experiments used 48 hours or less time on one of the mentioned hardware. 

\subsection{DSM pretraining}

\begin{wrapfigure}{l}{0.25\textwidth}
  \centering
  \includegraphics[width=\linewidth]{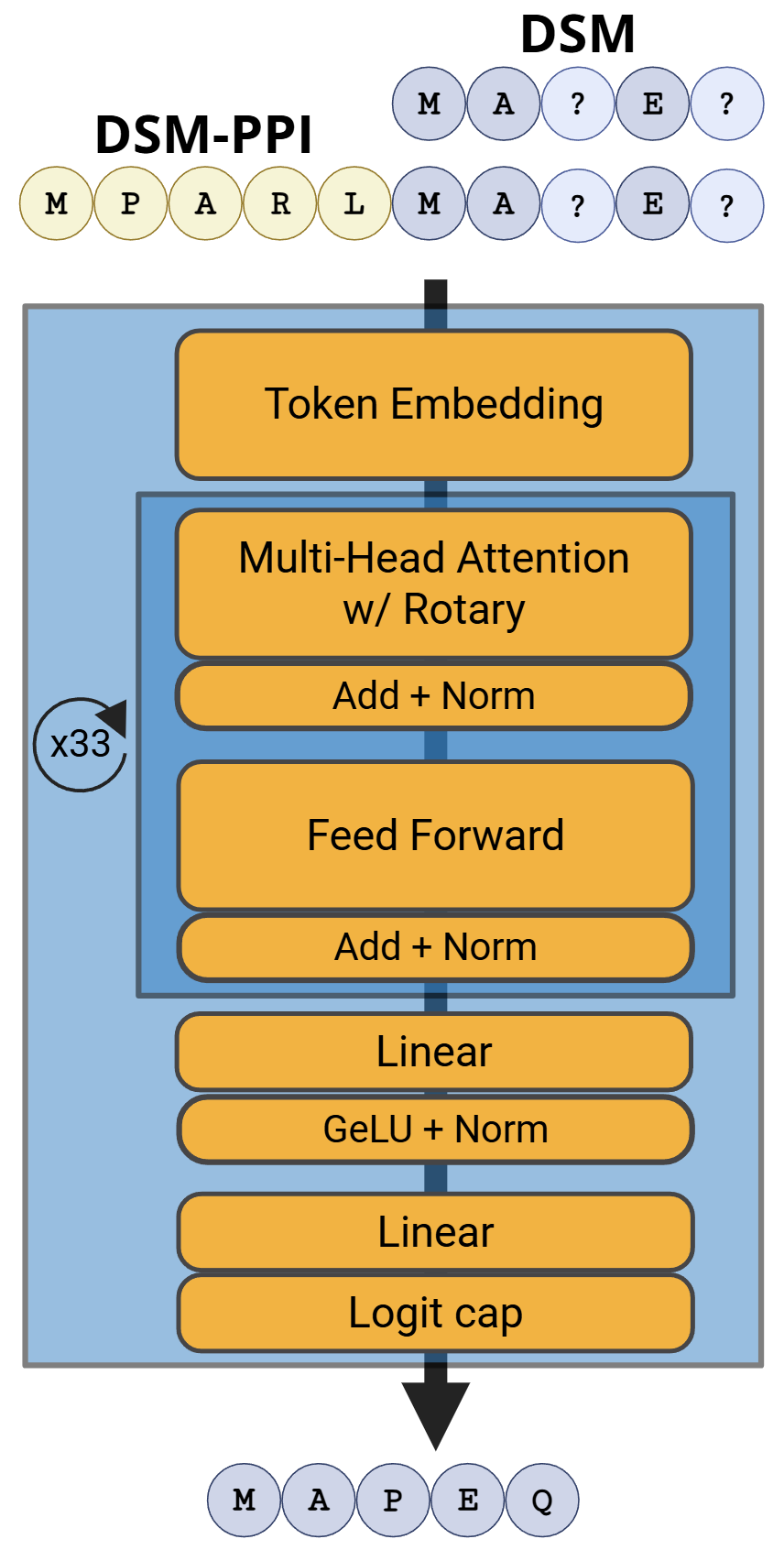}
  \caption{Model architecture with DSM and DSM-PPI training scheme highlighted. Amino acids are masked uniformly and filled in completely with each forward pass, either with or without interacting protein context.}
  \label{fig:model_arc}
\end{wrapfigure}

For DSM pretaining, we used OMG$_{prot50}$, a protein-only dataset containing over 207 million sequences after clustering the Open MetaGenomic dataset (OMG) at 50\% sequence identity \cite{tattabio_glm2_omg}. We randomly removed 10,000 sequences for the validation and test sets each. We queried UniProt (release 2024\_06) for newly deposited entries after 8/17/2024 (OMG release) with a minimum of transcript-level evidence. These 3,300 entries were added to the test set to accommodate for training sets with unknown composition before OMG use.

DSM models were extended from pretrained ESM2 checkpoints \cite{lin_evolutionary-scale_2023}. To stabilize training and improve sampling quality, we modified the language modeling head by adding an additional linear transformation as well as a soft-logit cap \cite{gemma2, soft_logit}. We tied the final projection to the token embedding matrix during training, which has been shown to be beneficial for representation learning \cite{deberta3, modernbert}.

During DSM training, we used a cosine learning rate scheduler with a 1,000-step linear warmup from zero \cite{cosine_lr}. We trained two versions of DSM, the first to prototype our process from the ESM2 150 million parameter checkpoint (ESM2$_{150}$) for 100,000 steps with a batch size of 32, maximum sequence length of 512, and learning rate of $1e^{-4}$ using the AdamW optimizer \cite{adamw}. This resulted in the \textbf{DSM$_{150}$} checkpoint. The larger version of DSM used the ESM2$_{650}$ checkpoint with the same training scheme of DSM$_{150}$, except with a global batch size of 128 and a maximum sequence length of 2048 to produce \textbf{DSM$_{650}$}. 

\subsection{Model probing}

Models were evaluated with a linear probe (\textbf{Supplemental Section \ref{sec:supp_Protify}}) after embedding supervised datasets (\textbf{Supplemental Section \ref{sec:supervised_data}}) with mean pooling over the last hidden state. Importantly, we gauged model performance against two controls. One was a negative control, gathered by replacing the pLM embeddings with randomly generated vectors. The second control was based on an ESM2$_{35M}$ copy that has randomized weights. We previously established that running similar inputs through a randomized transformer would still place them relatively close in the final latent space \cite{AV, random_transformers}, presumably due to the self-attention mechanism attending to similar regions of input tokens regardless of the starting values of the token embeddings. This is particularly powerful for pLMs, where protein sequences share a lot of functional relevance based on sequence homology. So we viewed the \texttt{Random Transformer} control as a benchmark for the correlation based on basic sequence homology. To easily view the gain over the negative control metric $r$, we linearly normalized the current score $p$ against the best score $q$ by calculating $\bar{p} = \frac{p-r}{q-r}$, where 1.0 is the best performer (\textbf{Figure \ref{fig:norm_rep}}, raw scores in \textbf{Supplemental Figure \ref{fig:supp_rep}}). In total, we benchmarked DSM against ESM2, GLM2, ProtBert, ProtCLM$_{1B}$ (an AR pLM from Biomap), ESMC (ESM++), DPLM, ANKH, and ProtT5 \cite{lin_evolutionary-scale_2023, tattabio_glm2_omg, elnaggar_prottrans_2022, cheng_training_2024, esm2024cambrian, ESM++, bytedance_dplm, elnaggar_ankh_2023}.

\subsubsection{Secondary structure prediction}

Secondary structure performance was screened as a token-wise multiclass classification problem using the same training schema as the linear probes, but with a single block transformer that utilized full-residue embeddings instead. We chose a hidden size of 512 for the vanilla transformer block with rotary embeddings \cite{vaswani_attention_2017, rotary_emb}. The best-performing base pLM was fully fine-tuned with the same training parameters, except with low rank adaptation (LoRA) applied to the attention layers instead of an external probe \cite{LORA}.

\subsection{Distribution comparisons}
\label{sec:dist}

To analyze and contrast the distributions of amino acids, secondary structures, and AV terms, we used a $\chi^2$ test for independence alongside the Jensen-Shannon divergence (JS) \cite{Pearson_1900, js_paper}. For amino acids and secondary structures, we independently compared the 1-mers, 2-mers, and 3-mers. For AV, we used only 1-mers due to the large vocabulary size. We hypothesized that \textit{de novo} generated proteins would inhabit distributions that reject the $\chi^2$ null due to the high sample count and degrees of freedom, but showcase a low JS. We believed this would imply biologically mimetic but distinct distributions given low $\chi^2$ $p$-values yet high similarity implied from JS. For $\chi^2$, we used the raw counts, and for calculating JS we converted to frequencies after additive smoothing to avoid division by zero \cite{additive_smoothing}.

\subsection{Unconditional generation analysis}

Unconditional generation of DSM models was accomplished by feeding in only mask tokens of a chosen length surrounded by \texttt{[CLS]} and \texttt{[EOS]}. This procedure simulates a reverse diffusion process via progressive denoising. The total number of forward steps was calculated based on the number of tokens and a step divisor $s$, with approximately $s$ tokens unmasked each step. Technically, every forward pass should fill in every mask after pretraining, so $s$ tokens were chosen to keep, and the rest were remasked. There are many possible strategies to choose which $s$ tokens to keep, such as logit or confidence-based ranking, topk, and beam search, etc. We observed in local experiments that pLMs are poorly conditioned to choose which position should be unmasked next; therefore, we used a random choice approach without search. To balance speed and quality, we swept over values of $s$ and sampling temperatures, optimizing for the lowest JS between the validation set and generated amino acid 3-mer distributions.

To analyze the trends of DSM generation, we chose the 10,000 validation sequences as a reference for natural proteins. A \textit{de novo} corpus was generated by looping through the validation set and generating a protein unconditionally based on the length of the protein in the validation set. The result was two corpora of amino acid sequences of exactly the same length, 9,989 total after removing entries less than 20 or greater than 2,048 long. Then, we applied the distribution comparisons mentioned above (\textbf{Section \ref{sec:dist}}).

\subsection{Sequence reconstruction evaluation}
\label{sec:seq_recon}
We define sequence reconstruction as the ability for a pLM to reconstruct noised inputs at various corruption levels. To evaluate DSM on sequence reconstruction, we masked the pretraining validation and test sets with 5\%, 15\%, 30\%, 50\%, 70\%, and 90\% mask rates, then recorded the cross-entropy loss and weighted F1 score over the masked positions and ground truth, as well as ASc. For comparison we did the same for ESM2$_{8}$, ESM2$_{35}$, ESM2$_{150}$, ESM2$_{650}$, and ESM2 three billion (ESM2$_{3B}$). Importantly, the same random seeds ensured that identical masked positions across models were used for each run, allowing for a fair comparison.

\subsection{DSM binder generation}

A new Protein-Protein Interaction (PPI) dataset was compiled by downloading the StringDB version 12 entries for model organisms, including \textit{Homo sapiens}, \textit{Mus musculus}, \textit{Arabidopsis thaliana}, \textit{Saccharomyces cerevisiae}, \textit{Drosophila melanogaster}, \textit{Danio rerio}, \textit{Caenorhabditis elegans}, \textit{Escherichia coli} (K12), and \textit{Pseudomonas aeruginosa }(PAO1) \cite{string_db}. Entries with a combined score less than 900 (90\% confidence) were discarded, and the set of unique sequences was put aside. CD-hit was run on the sequence set, clustering at 90\% sequence identity, and the representative entries were saved \cite{cdhit}. We only retained PPI examples from our compiled list where both sequences were representatives. We split a random section of the dataset for evaluation and removed sequences matching an example in the training set. The final dataset consisted of 646,000 high-quality PPI entries for training, a random split of 5,850 entries for validation, and a data-leakage-free test set of 1,320 entries.

After data compilation and pretraining, we fine-tuned the DSM models to generate a candidate interacting sequence (SeqB) conditioned on a fixed target sequence (SeqA). We reasoned that most protein interactions take place through some type of inter-protein binding, so this training process simulated the task of \textit{de novo} binder design. This was achieved by masking the second sequence in the PPI entry. The format \texttt{[CLS]--SeqA--[EOS]--[MASKED~SeqB]--[EOS]} was used for the same exact objective as pretraining. The DSM$_{150-ppi}$ and DSM$_{650-ppi}$ were trained with LoRA on the attention layers for one epoch of the PPI dataset \cite{LORA}. A LoRA $r$ of 8, $\alpha$ of 32, and dropout of 0.01 were used. We trained a control version of DSM$_{150-ppi}$ with the same exact hyperparameters but with inputs \texttt{[CLS]--SeqB--[EOS]} only. The control model omits conditioning on SeqA, allowing us to confirm that improvements are due to context-awareness rather than simple additional fine-tuning. Importantly, the order of SeqA and SeqB was switched randomly throughout training, but not testing.

We confirmed the ability for DSM to utilize SeqA context to get better at reconstructing SeqB by applying the sequence reconstruction pipeline (\ref{sec:seq_recon}) of ESM2 and DSM models on the PPI test set at a 15\% mask rate. ESM2 and base DSM models received only SeqB, DSM$_{150-ppi-control}$ also received only SeqB, and the DSM$_{ppi}$ versions received both sequences.

We established a pipeline for template-based design in an attempt to use DSM to create stronger binders. To explore well-known target-binder pairs, we conducted an extensive literature review for the targets in the brand-new BenchBB protein binder benchmark \cite{egfr_comp}. BenchBB is a compilation of standardized protein targets (EGFR, IL-7R$\alpha$, MBP, PD-L1, BBF-14, BHRF1, Cas9) for rigorous and consistent evaluations in computational protein binder design. We sourced their current best publicly available binders, described in \textbf{Supplemental section \ref{sec:lit_review}}.

To explore the relevant sequence space near the known binder templates, we conducted large-scale screens based on random masking. With 100,000 iterations for each protein target, we randomly masked its best-known binder between 0 and 100\% uniformly. Half of the time, we also uniformly sampled a random section of the template to use instead of the entire sequence in an attempt to hone in on domains vital to binding activity. This design enables the exploration of both global and localized binding-relevant motifs, promoting diversity while maintaining biological plausibility. After masking, the template sequence was fed to a DSM model with $s=100$ and a temperature of 1.0. Target proteins and their newly designed binders were sent to Synteract2 through the Synthyra API to predict their binding affinity \textbf{Supplemental section \ref{sec:supp_Synteract2}} \cite{hallee_protein-protein_2023}. \texttt{Unconditional} generation refers to DSM$_{650}$ designing a version of the template \textit{without} reference to the target. \texttt{Conditional} generation refers to DSM$_{650-ppi}$ designing a version of the template \textit{while} referencing the target.

\section{Results and Discussion}
\label{sec:results}

\begin{wrapfigure}{r}{0.65\textwidth}
  \centering
    \begin{subfigure}{0.65\textwidth}
        \centering
        \includegraphics[width=\textwidth]{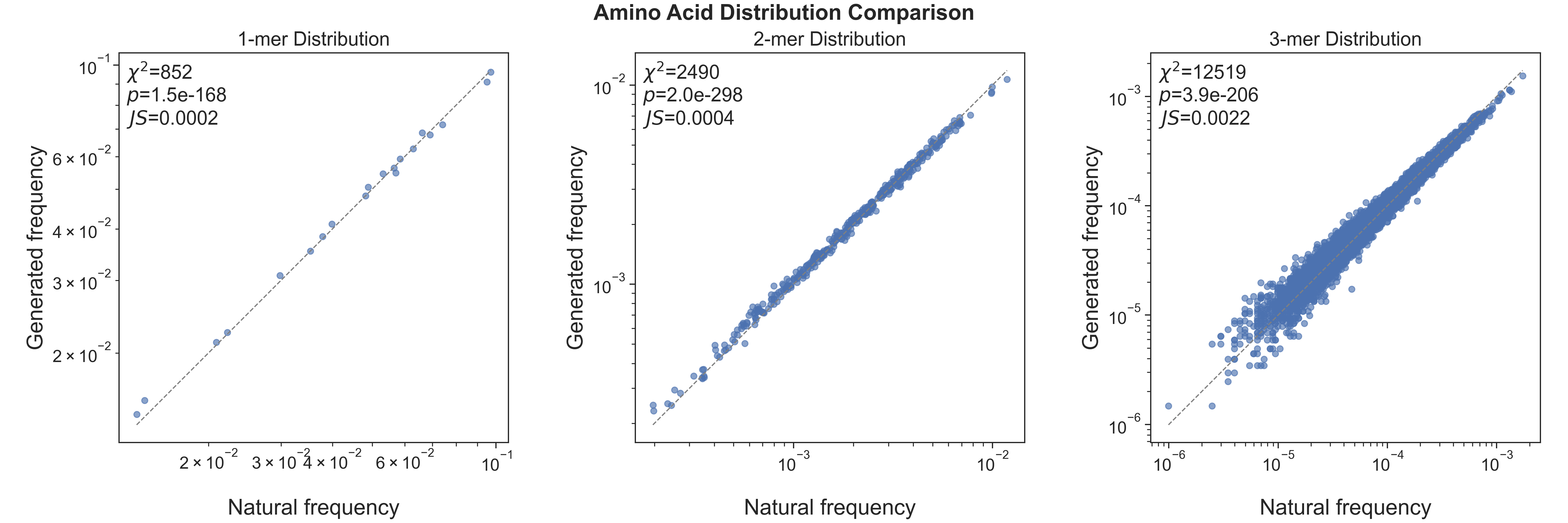}
    \end{subfigure}
    \vspace{0.5em}
    \begin{subfigure}{0.65\textwidth}
        \centering
        \includegraphics[width=\textwidth]{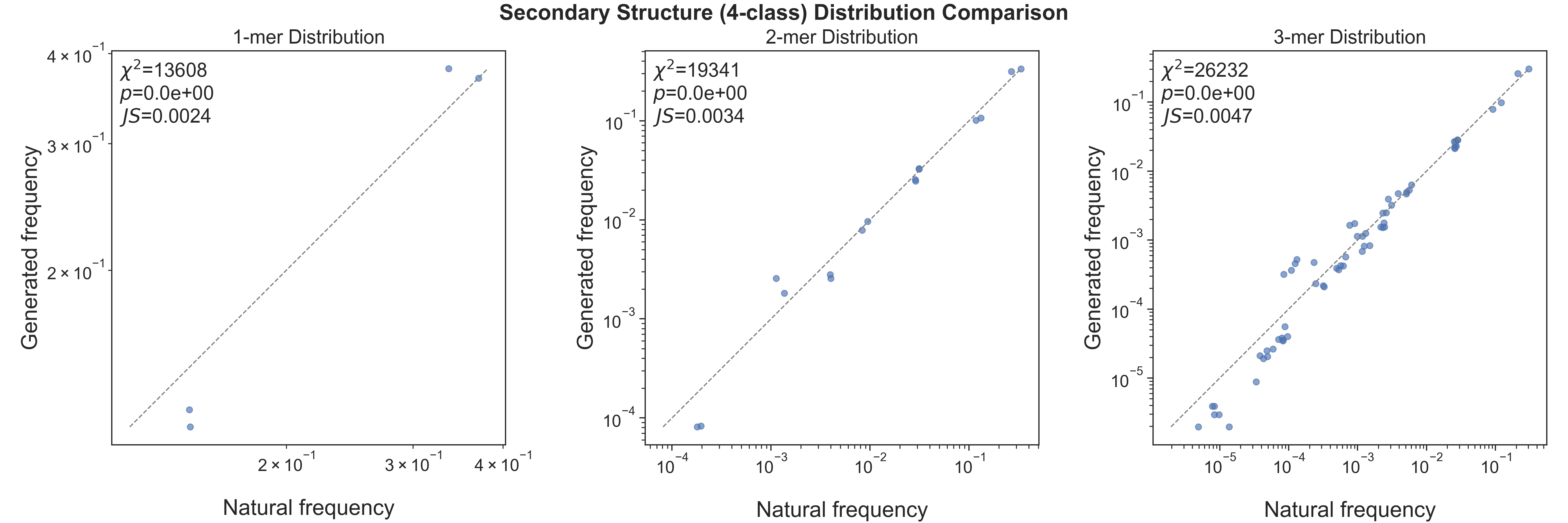}
    \end{subfigure}
    \vspace{0.5em}
    \begin{subfigure}{0.65\textwidth}
        \centering
        \includegraphics[width=\textwidth]{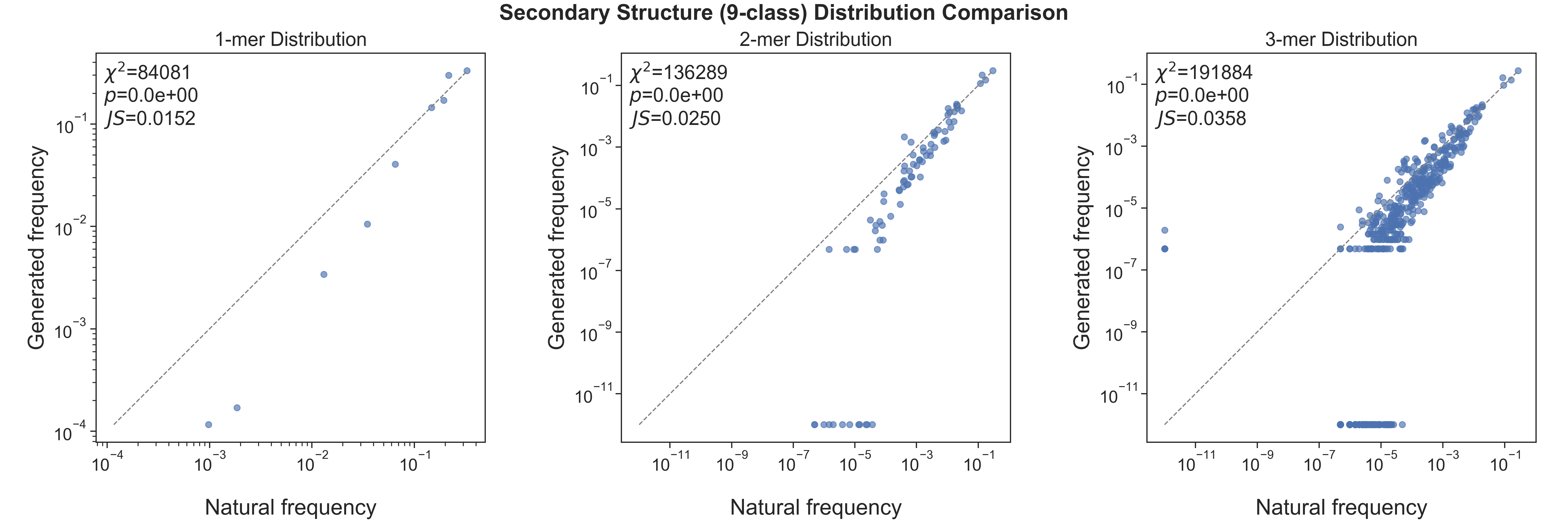}
    \end{subfigure}
  
  \caption{Plot of natural vs. generated 1-mer, 2-mer, and 3-mer comparisons with associated $\chi^2$ values and JS. Amino acid frequencies (top row), secondary structure [four-class] token frequencies (middle row), and secondary structure [nine-class] token frequencies (bottom row).}

  \label{fig:dist}
\end{wrapfigure}

\subsection{DSM generated biomimetic but distinct sequence distributions}
We conducted a grid search over sampling temperature and $s$, identifying that temperature = 1.0, no gumbel softmax, and filling in one token at a time ($s=1$) yielded the lowest JS with natural sequences. However, we observed that the JS of amino acid 3-mers was almost unchanged between $s=1$ and $s=5$, so we chose $s=5$ for future experimentation, resulting in five times faster throughput. After the hyperparameter search, we generated ~10,000 sequences unconditionally based on the validation set. Then, we predicted the secondary structure and AV terms for the natural and generated sequences, leaving us with three types of distributions to compare between natural and our \textit{de novo} proteins. While the $\chi^2$ test rejects the null hypothesis of identical distributions (likely due to large sample size), the low JS indicates high similarity.

When comparing natural vs. generated frequencies, we expected entries to fall close to $y=x$. Indeed, we saw that the amino acid k-mers followed $y=x$ closely with very low JS values $< 0.01$, yet low $\chi^2$ $p$-values (\textbf{Figure \ref{fig:dist}}). This demonstrates that DSM has a strong understanding of biological amino acid usage. The secondary structure comparison followed a similar trend, with low JS and low $p$-values (\textbf{Figure \ref{fig:dist}}). We reasoned that the strict alignment of the four-class secondary structure implied that DSM produced sequences with natural-like or convincing structural features. Lastly, the nine-class secondary structure followed similar trends (\textbf{Figure \ref{fig:dist}}). However, there is some clear nuance in the distribution of unconditional DSM outputs. Whereas a few secondary structure k-mers are overrepresented, the vast majority are underused. We viewed the nine-class 2-mers and 3-mers as fairly niche structural regions that may be linked to a highly specific biochemical function. With this in mind, these data imply that DSM produces `generic` possible sequences, at least without steering towards a specific type of protein.

\begin{figure}[t]
  \centering
  \begin{subfigure}[b]{0.28\textwidth}
    \includegraphics[width=\linewidth]{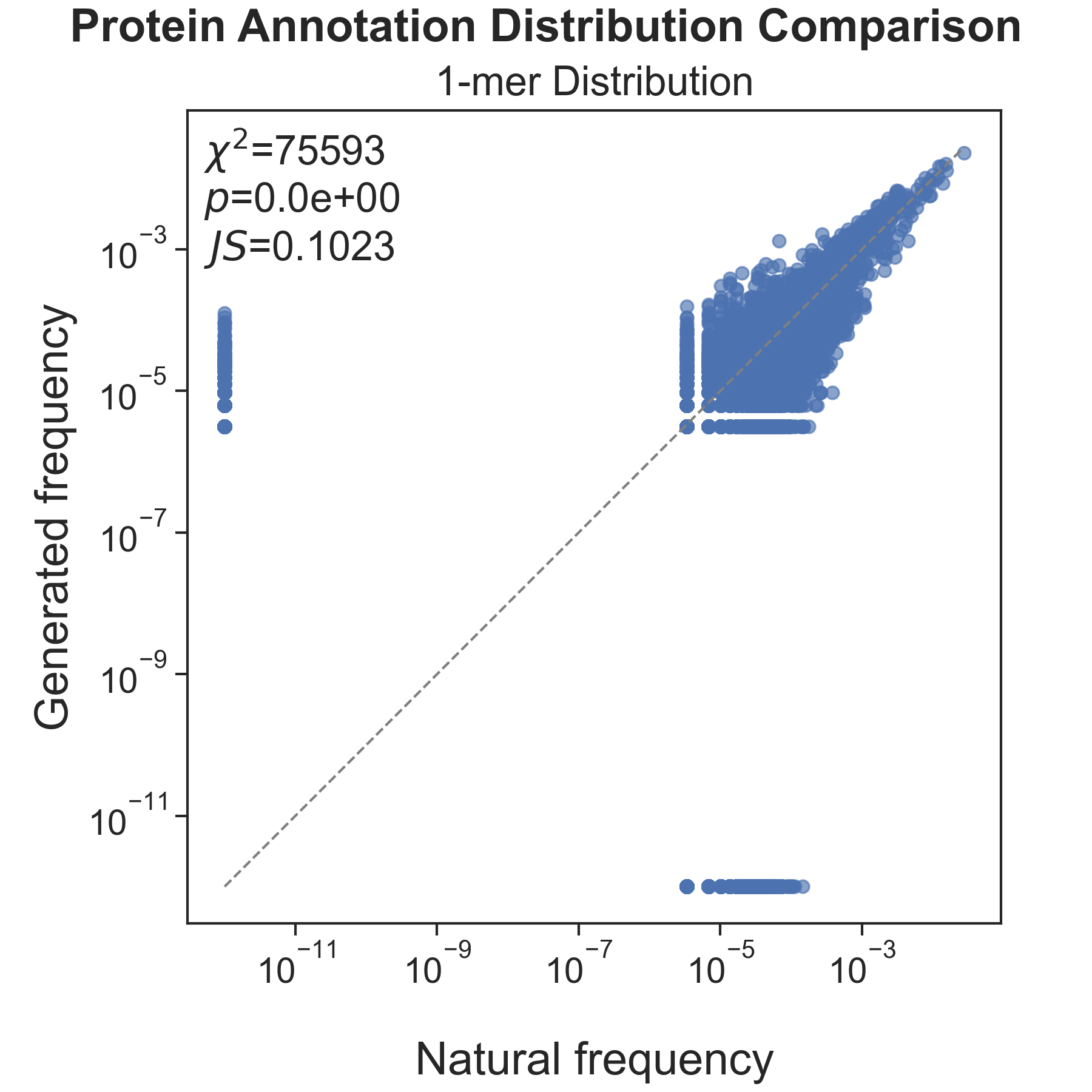}
  \end{subfigure}
  \hspace{1em}
  \begin{subfigure}[b]{0.62\textwidth}
    \includegraphics[width=\linewidth]{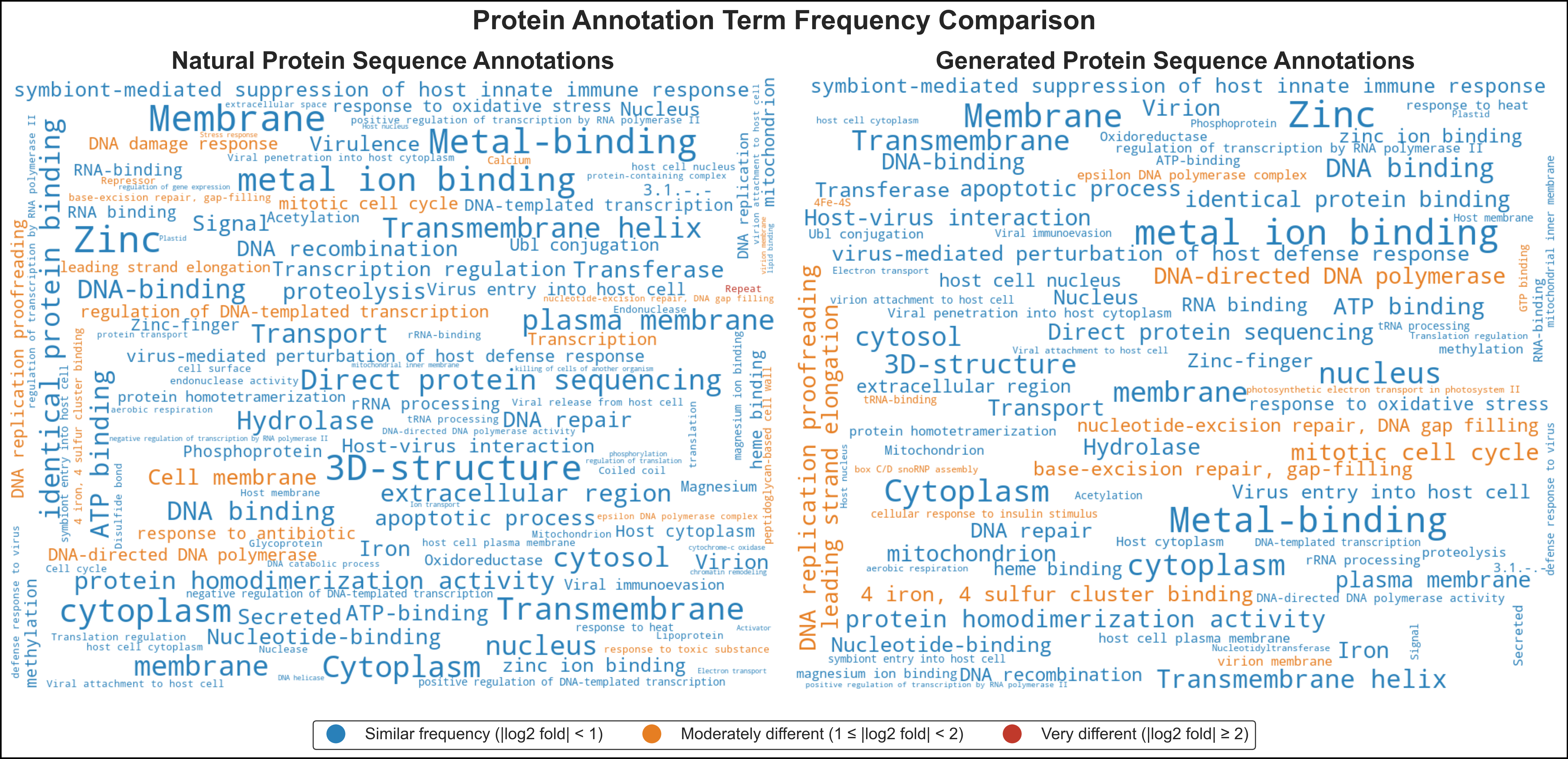}
  \end{subfigure}

  \caption{(Left) Plot of natural vs. generation AV token 1-mers with $\chi^2$ values and JS. (Right) Word cloud of natural and generated sequence annotations. Higher frequency terms have a bigger font, and terms are colored based on similarity between natural and generated frequencies via $\log_2$ fold change.}
  \label{fig:ann}
\end{figure}

When the AV terms were compared, we saw that natural and generated sequences were described by many unique AV terms, with the rest much closer to $y=x$. This resulted in a very low $p$-value but higher JS value of 0.102 (\textbf{Figure \ref{fig:ann}A}). Given that the AV vocabulary exceeds 88,000 terms, with some overlapping usage, we hypothesized that increased sequence sampling would further reduce JS. However, such increases in sample size would likely exacerbate $\chi^2$ test sensitivity, making statistical significance less informative. Closer inspection of the predicted functions shared between the natural and generated proteins (\textbf{Figure \ref{fig:ann}B}) demonstrated high overlap among common terms, combined with low $p$-values and JS across amino acid, secondary structure, and AV term k-mers. This suggests that DSM generates biomimetic sequences that occupy a distinct yet biologically plausible distribution when unconditional generation is used. This balance, capturing the latent structure of biochemical systems, may be ideal for fine-tuning towards specific design tasks.

The secondary structure model screening (\textbf{Supplemental Figure \ref{fig:supp_ss_screen}}) and final production model performance (\textbf{Supplemental Figure \ref{fig:supp_ss_prod}}) along with protein annotation performance (\textbf{Supplemental Section \ref{sec:supp_Translator}}) can be explored in the Supplemental Material.

\subsection{DSM demonstrated substantially improved sequence reconstruction compared to MLM counterparts}

\begin{figure}[b]
  \centering
  \includegraphics[width=\textwidth]{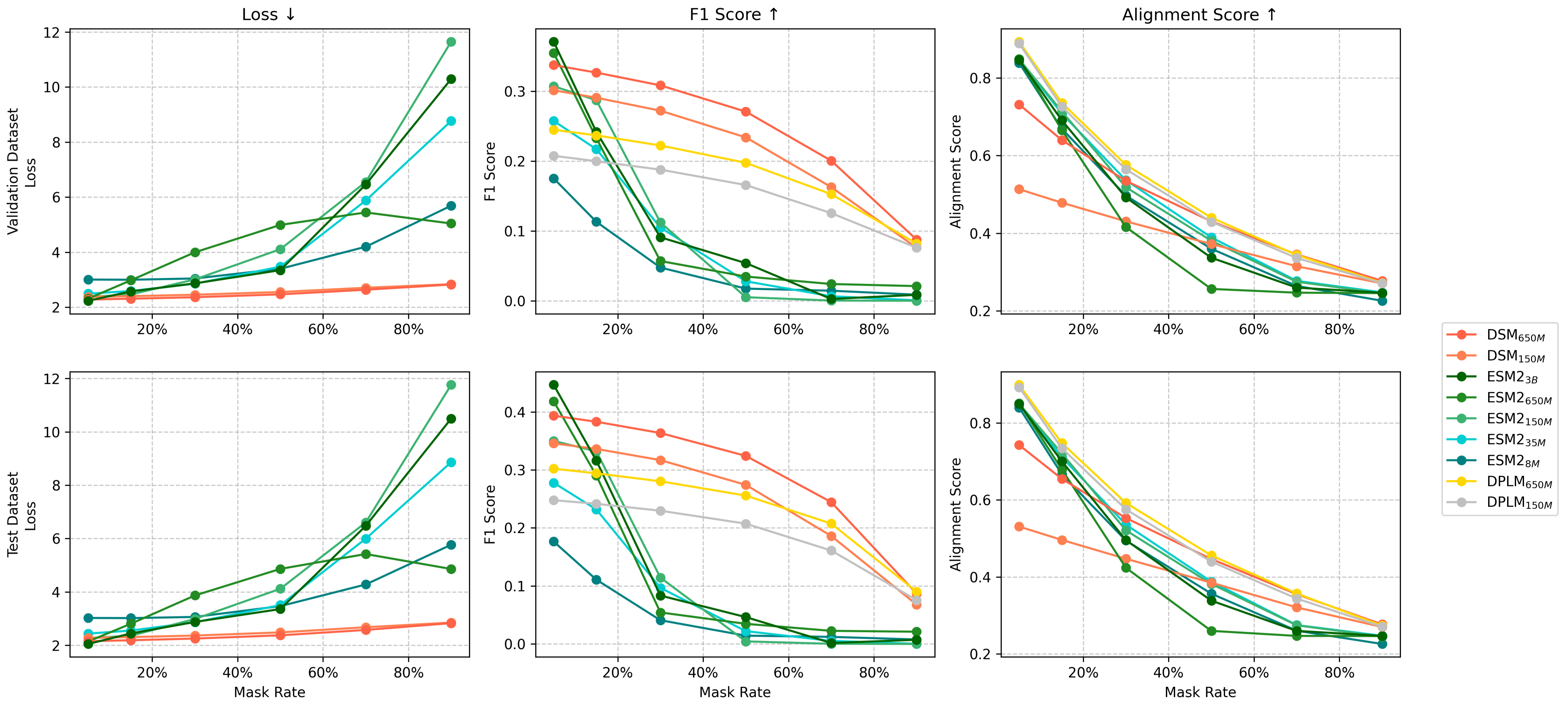}
  \caption{Loss, weighted F1, and ASc for DSM, DPLM, and ESM2 models across a range of mask percentages for the validation (top row) and test sets (bottom row).}
  \label{fig:mask_fill}
\end{figure}
In line with more direct comparisons to existing MLM-based pLMs, we examined sequence reconstruction with various mask rates, filling in 
masked sequences with a single forward pass. We evaluated DSM against MLM-based ESM2 models, ranging from eight million to three billion 
parameters, as well as DPLM models, which employ a discrete diffusion process. Of note, the DPLM \texttt{generate} function does not output raw logits, so they were omitted from the loss analysis.

We observed that DSM models exhibited worse performance compared to ESM2 models at low mask percentages (5-15\%); however, across higher mask rates, they maintained a low loss and high reconstruction metrics compared to MLM-only training (\textbf{Figure \ref{fig:mask_fill}}). While the DPLM models had similar performance to DSM models, even matching alignment scores at high mask rates, DSM models showed a considerable gain in the F1 scores, ranging from 2.4\% to 37.8\% higher based on the mask percentage.

Interestingly, the diffusion-based pLMs showcased a distinct sequence reconstruction scaling law that became clear when contrasting DSM and DPLM compared to ESM2 (\textbf{Figure \ref{fig:mask_fill}}). Even at 90\% masking, the diffusion models produced highly similar sequences to the ground truth with an ASc \textasciitilde{0.27}; DSM slightly outperformed DPLM (0.2771 vs. 0.2721 ASc on validation, 0.2766 vs. 0.2734 ASc on test). Both models demonstrated impressive reconstruction ability, as 0.27 ASc is over four standard deviations above the mean of randomly paired natural protein sequences. Importantly, these sequence reconstruction metrics show that DSM could generate realistic protein sequences from a \textit{single} forward pass. This efficiency positions DSM as a compelling alternative to AR or discrete diffusion models, especially in large-scale or real-time design settings.

\subsection{DSM produced high-quality embeddings in addition to generative capabilities}

\begin{figure}[b]
  \centering
  \includegraphics[width=\textwidth]{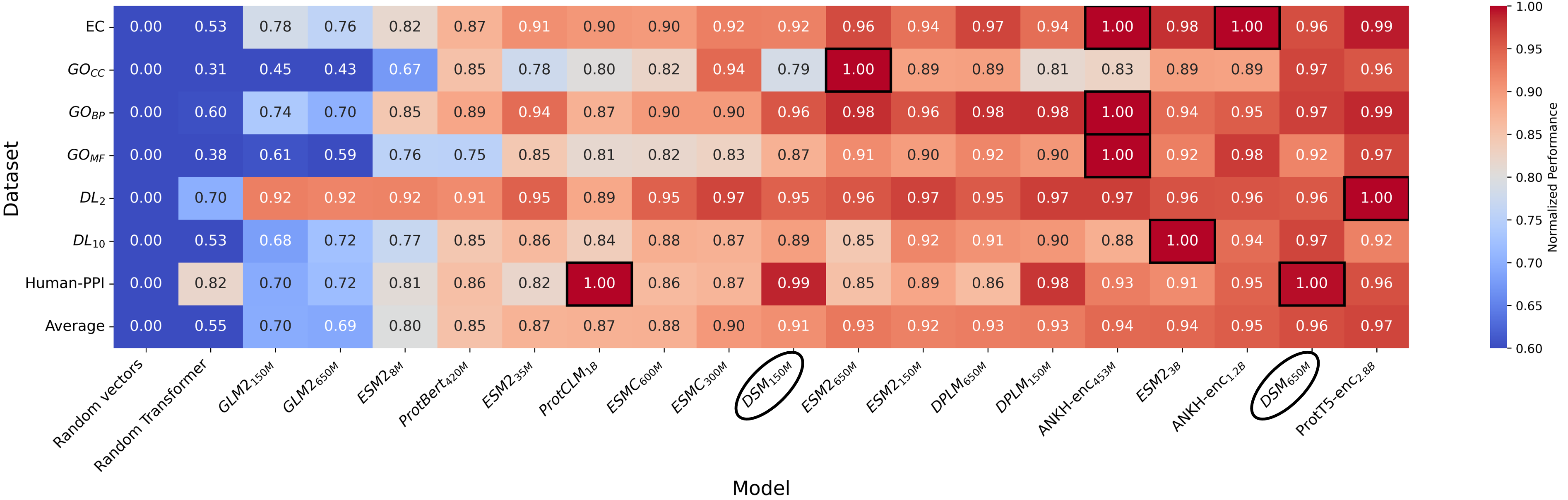}
  \caption{Protein representation probe scores, reported and colored by their relative F1 score increase over the random vector control.}
  \label{fig:norm_rep}
\end{figure}

Another important use case for pLMs is protein annotation based on direct supervised learning or vector search. We chose to evaluate the propensity for annotation by probing frozen versions of the pLMs and assessing the intrinsic correlation between the embeddings and valuable downstream tasks (datasets described in \textbf{Supplemental Figure \ref{fig:supp_data}}). DSM$_{650}$ produced the highest quality embeddings among similarly sized pLMs, generating consistently high F1 scores across a wide variety of tasks, only overtaken by the much larger ProtT5 on average (\textbf{Figure \ref{fig:norm_rep}}). Importantly, we saw a boost in performance over its base weights of ESM2$_{650}$, although all the model performances were much better than random and similar (\textbf{Supplemental Figure \ref{fig:supp_rep}}). We observed a striking difference between the autoregressive pLM tested, ProtCLM, and the diffusion-based pLMs, DPLM and DSM, suggesting that diffusion offers a promising approach to unify generative capabilities and representation quality.

\begin{figure}[t]
  \centering
  \includegraphics[width=.85\textwidth]{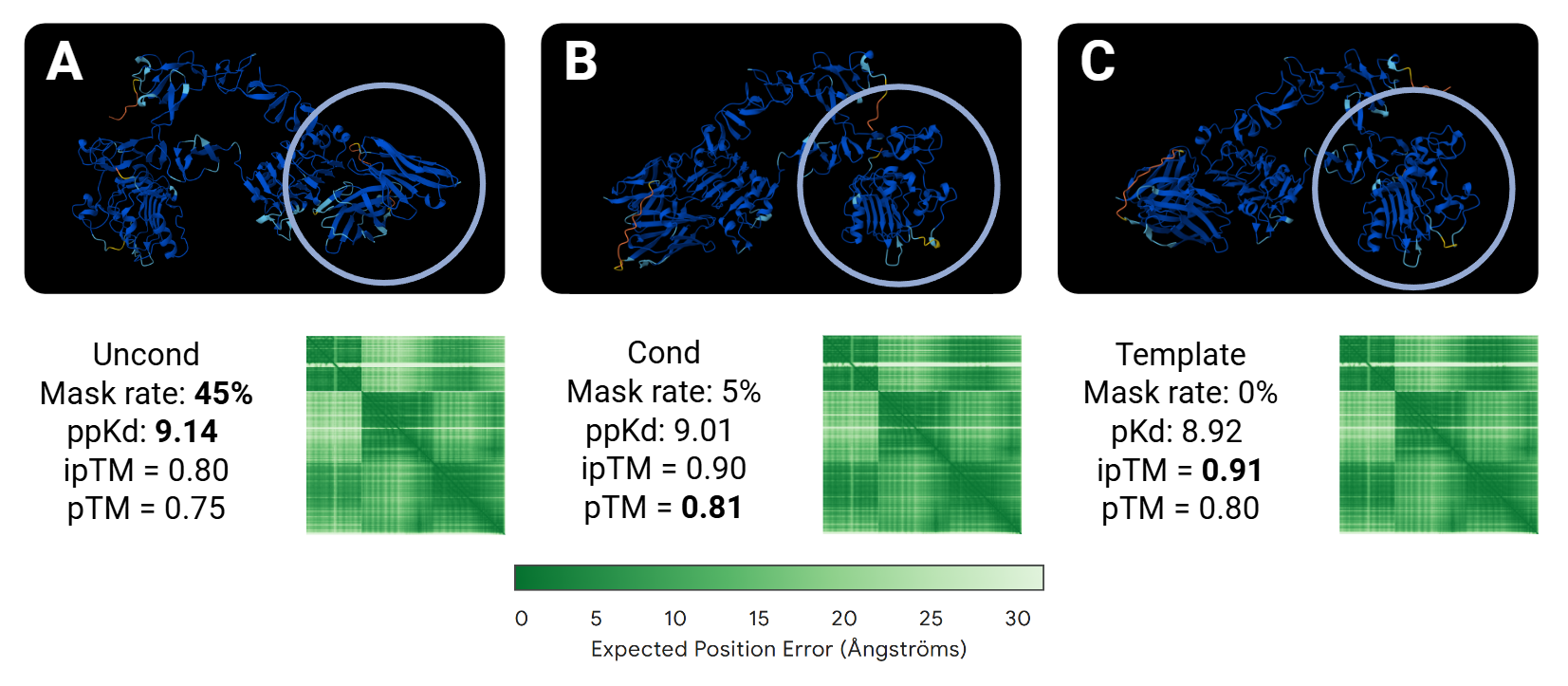}
  \caption{Designed EGFR binders folded alongside EGFR using AlphaFold3, associated plots from web server rendered \cite{af3}. (A) Unconditional generation at a high mask rate with a very high ppKd and compliance with AlphaFold3. (B) Conditional generation with the highest recorded AlphaFold3 pTM of folded samples from either method. (C) Adapytv competition winner with the highest ipTM of folded designs \cite{egfr_comp}}.
  \label{fig:egfr_highlight}
\end{figure}

\subsection{DSM can be fine-tuned to design protein binders}

We confirmed that DSM models could successfully use interacting protein data to reconstruct sequences. Whereas DSM$_{150-ppi-control}$ exhibited a small increase in reconstruction accuracy compared to DSM$_{150}$ (31.4\% $\rightarrow$ 31.9\%), DSM$_{150-ppi}$ had 33.0\% accuracy. We saw similar trends in all the reported metrics, and much lower losses between ESM2 and DSM base models relative to DSM$_{ppi}$ versions. In particular, DSM$_{650-ppi}$ achieved a cross-entropy loss of 1.989, the lowest among all tested models (full metrics in \textbf{Supplemental Table \ref{tab:supp_ppi}}).

To apply these generative abilities, we generated 1.4 million potential interactors, 100,000 for each BenchBB target and template-guided scheme, with the goal of producing proteins with high binding affinities. Interestingly, predicted pKd (ppKd) values remained stable across mask rates, suggesting DSM's robustness to input corruption in template-based binder design. Top binder designs by ppKd for each target and method can be found in \textbf{Supplemental Figure \ref{fig:supp_best_binders}}; the success rates and average ppKd for each target-method combination can be found in \textbf{Supplemental Table \ref{tab:supp_benchbb}}. Some compelling EGFR binders (\textbf{Figure \ref{fig:egfr_highlight}}) showcase two notable designs folded alongside EGFR using AlphaFold3 \cite{af3}. One design (\textbf{Figure \ref{fig:egfr_highlight}A}) exhibits higher predicted binding affinity (ppKd) than the best known binder (\textbf{Figure \ref{fig:egfr_highlight}C}) from the EGFR competition while maintaining high ipTM and pTM scores, despite having 45\% of the original template hidden from DSM. The protein in \textbf{Figure \ref{fig:egfr_highlight}B} highlights the only folded design that exhibits a higher pTM compared to the original template, while also having a higher ppKd.

\section{Conclusion}
\label{sec:conclusion}

We introduced Diffusion Sequence Modeling (DSM), a simple yet powerful way to retrofit any MLM-based pLM with a diffusion objective. With only minor modifications to the masking scheme, loss calculation, and custom logit-cap head, DSM turns transformer encoders into protein designers. DSM achieves strong sequence reconstruction performance on unseen data across high corruption levels and generates biomimetic proteins that capture the statistical and functional properties of natural sequences. Additionally, on diverse representation learning benchmarks, DSM outperforms state-of-the-art MLM, autoregressive, and discrete diffusion pLMs. Fine-tuning DSM for conditional binder generation further enables template-guided design of interacting proteins, showcasing strong results on the new BenchBB benchmark.

While our study relied on \textit{in-silico} proxies for secondary structure, function, and binding affinity, the framework is agnostic to the supervision source; integrating wet-lab feedback or structure-aware losses from crystal structures is a natural next step. We also envision DSM variants for mutagenesis studies, multimodal conditioning (including structure, small molecules, text, and AV tokens), and active learning loops to iteratively refine the model \cite{su_saprot_2023, protokens, AV}. Other biological sequence modalities, such as codons, could further refine organism-specific protein production optimization through codon usage refinement \cite{hallee_cdsbert_2023, hallee_machine_2023}. We also believe architectural changes, for example, mixture-of-experts systems with modality-based routing, could also improve the embedding and generation quality \cite{hallee_moe_2025}.

It is important to note that pLMs have the potential to revolutionize numerous economic sectors, including healthcare and environmental sciences, while also carrying associated risks. In our work, we proactively mitigate unintended risks by excluding disease-related annotation terms and actively avoid training models on proxies for virulence, with the intention of not directly enabling types of bioterrorism. This approach aligns with broader concerns in the scientific community about the dual-use nature of AI in biology, where models could potentially lower barriers for malicious actors designing harmful biological entities \cite{responsible_ai}.

In summary, DSM closes the gap between \textit{understanding} and \textit{creating} proteins in a single architecture, enabled by a few Python line changes, providing a foundation for rapid, scalable, and biologically grounded protein design.

\vspace{1em}
\begin{dataavail}\noindent
Selected datasets, code, and model weights can be found at \href{https://github.com/Gleghorn-Lab/DSM}{github.com/Gleghorn-Lab/DSM}.
\end{dataavail}

\begin{interests}\noindent
LH, DB, and JPG are co-founders of and have an equity stake in Synthyra, PBLLC.
\end{interests}

\begin{contributions}\noindent
Conceptualization (LH, JPG), Synthyra API (LH, DB), Model architectures (LH), Data Curation (LH), Investigation (LH, NR), Formal Analysis (LH, NR, JPG), Writing – Original Draft (LH, JPG), Writing – Review \& Editing (LH, NR, DB, JPG), Supervision (JPG), Project Administration (JPG), Funding acquisition (LH, JPG).
\end{contributions}

\begin{acknowledgements}\noindent
The authors thank Katherine M. Nelson, Ph.D., for reviewing and commenting on drafts of the manuscript. This work was partly supported by the University of Delaware Graduate College through the Unidel Distinguished Graduate Scholar Award (LH), the National Science Foundation through NAIRR pilot 240064 (JPG), and the National Institutes of Health through NIGMS T32GM142603 (NR), R01HL133163 (JPG) and R01HL145147 (JPG).
\end{acknowledgements}

\section{References}
\bibliography{references}

\begin{thebibliography}{76}
\bibitem[Pearson(1900)]{Pearson_1900}
Karl Pearson.
\newblock X. on the criterion that a given system of deviations from the probable in the case of a correlated system of variables is such that it can be reasonably supposed to have arisen from random sampling, July 1900.
\end{thebibliography}

\onecolumn
\newpage

\appendix

\section{Supplementary Material}

\renewcommand{\thefigure}{S\arabic{figure}}
\setcounter{figure}{0}
\renewcommand{\thetable}{S\arabic{table}}
\setcounter{table}{0}

\subsection{Synthyra models} 

Synthyra is a Public Benefit Company (PBLLC) that offers protein annotation services with deep learning models. While the Synthyra models used throughout this study are closed-source, anyone can use free credits to query the models at \url{https://Synthyra.com}. Here, we present various implementation details and performance metrics to further inform the results of this study.  

\subsubsection{Synteract2}
\label{sec:supp_Synteract2}
\begin{figure}[bp]
  \centering

  \begin{subfigure}[t]{.75\textwidth}
    \centering
    \caption{}
    \vspace{1em}
    \includegraphics[width=\textwidth]{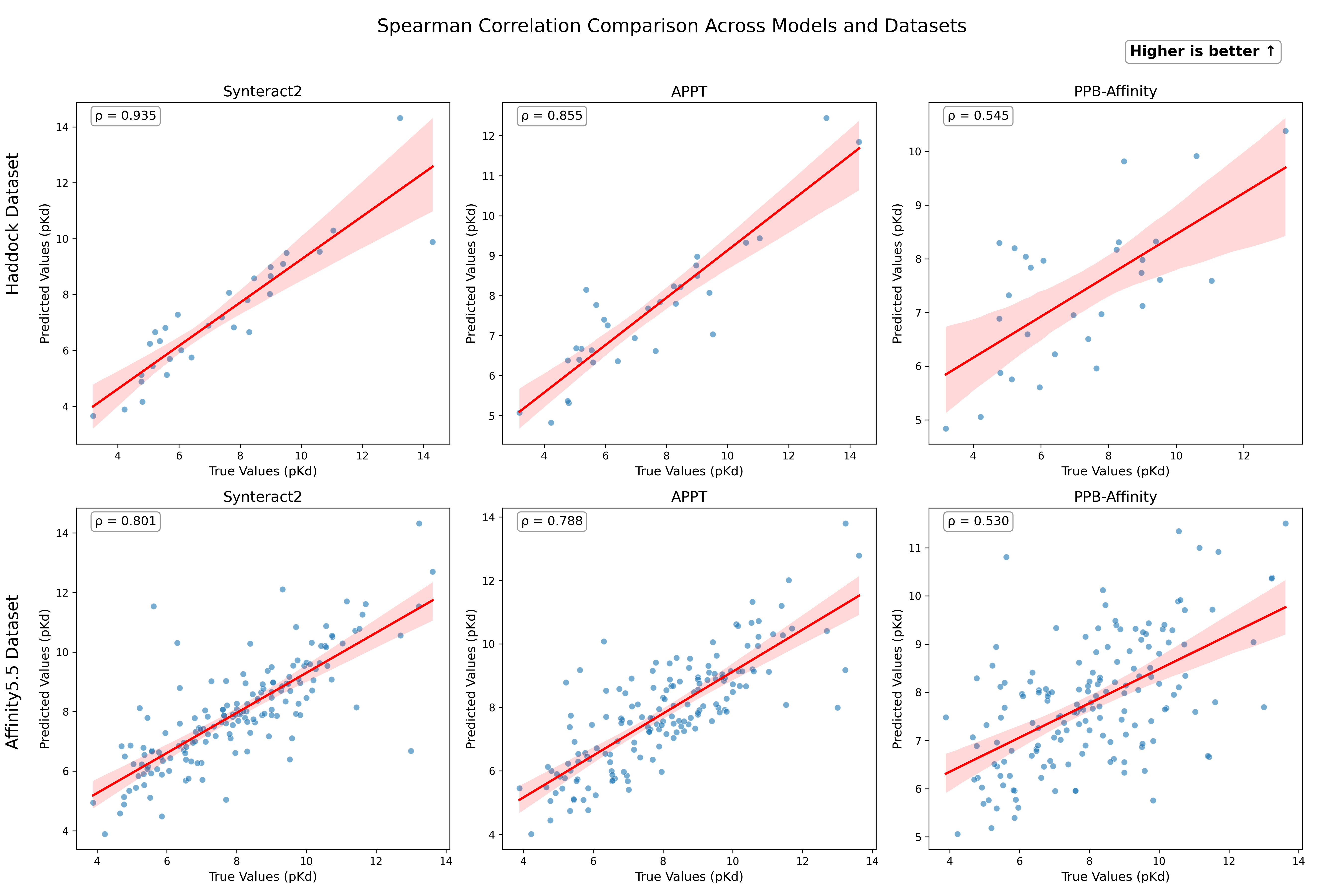}
    
  \end{subfigure}

  \vspace{1em}

  \begin{subfigure}[t]{0.65\textwidth}
    \centering
    \caption{}
    \vspace{1em}
    \includegraphics[width=\textwidth]{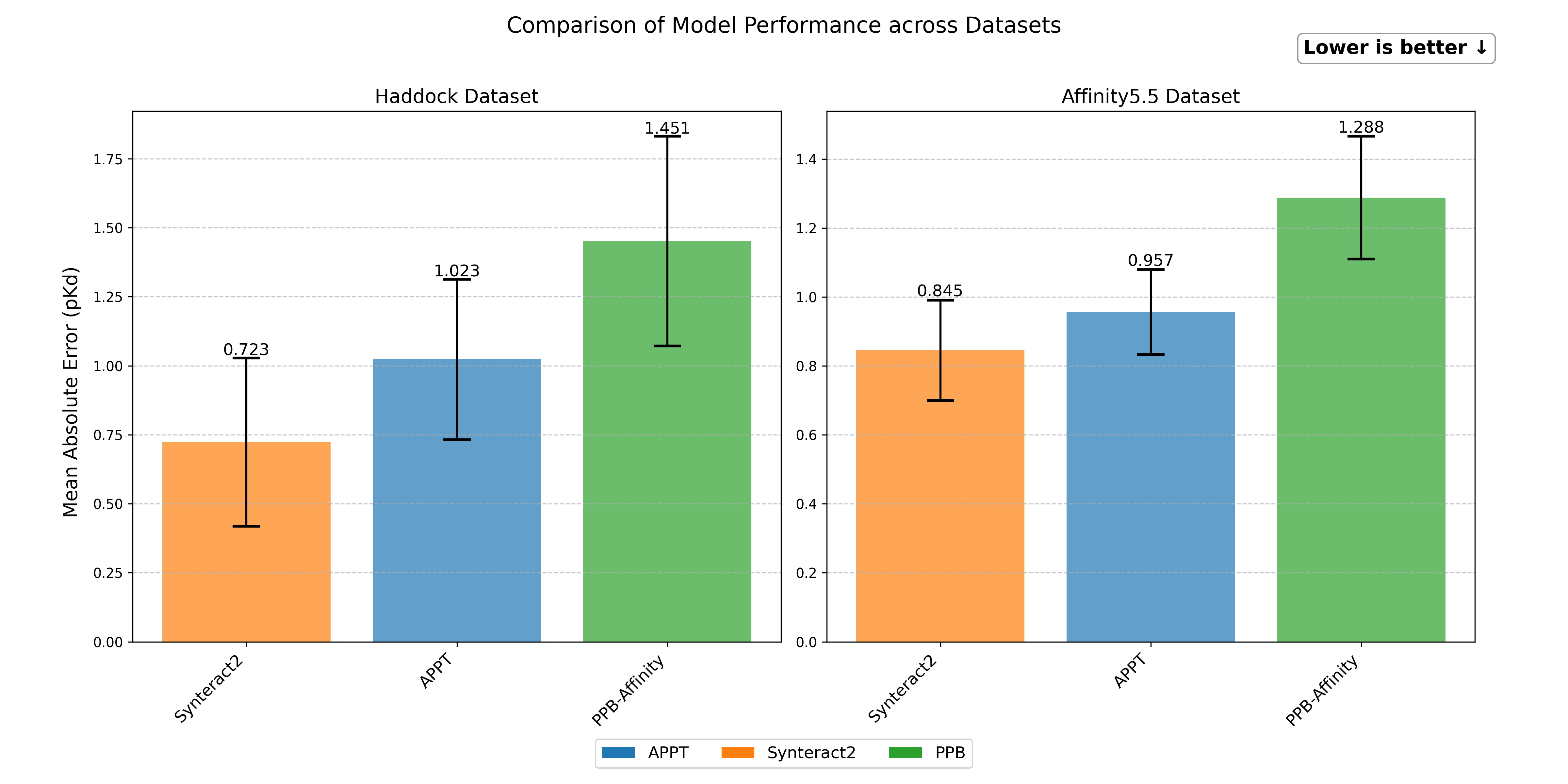}
    
  \end{subfigure}

  \caption{Predicted binding affinity performance, comparing Synteract2, APPT, and PDB-affinity on the Haddock and Affinityv5.5 benchmarks. (a) Spearman $\rho$ reported with line of best fit and 95\% confidence intervals are calculated with \texttt{seaborn regplot}, which employs bootstrapping \cite{Waskom2021}. (b) Mean absolute error with 95\% confidence intervals calculated via t-test.}
  \label{fig:supp_pkd}
\end{figure}

Synteract2 is the second generation in a line of LLMs built to model protein-protein interactions (PPI). Synteract2 jointly models the probability of PPI, plausible binding sites, and the binding affinity (pKd) between two input amino acid sequences. This study extensively used the binding affinity track of Synteract2 to screen potential designed binders. The binding affinity track of Synteract2 is extended from a BERT-like pLM trained with custom parameter-efficient fine-tuning methods and trained on the binding affinity dataset from the APPT project \url{https://github.com/Bindwell/APPT} - a processed version can be found here: \url{https://huggingface.co/datasets/Synthyra/ProteinProteinAffinity}. We highlight its leading performance on the Haddock and Affinityv5.5 benchmarks over previous state-of-the-art methods \textbf{ Supplemental Figure \ref{fig:supp_pkd}}.

\subsubsection{Translator}
\label{sec:supp_Translator}
\begin{wrapfigure}{r}{0.7\textwidth}
 % \centering

  \begin{subfigure}[t]{0.34\textwidth}
 %   \centering
    \includegraphics[width=\textwidth]{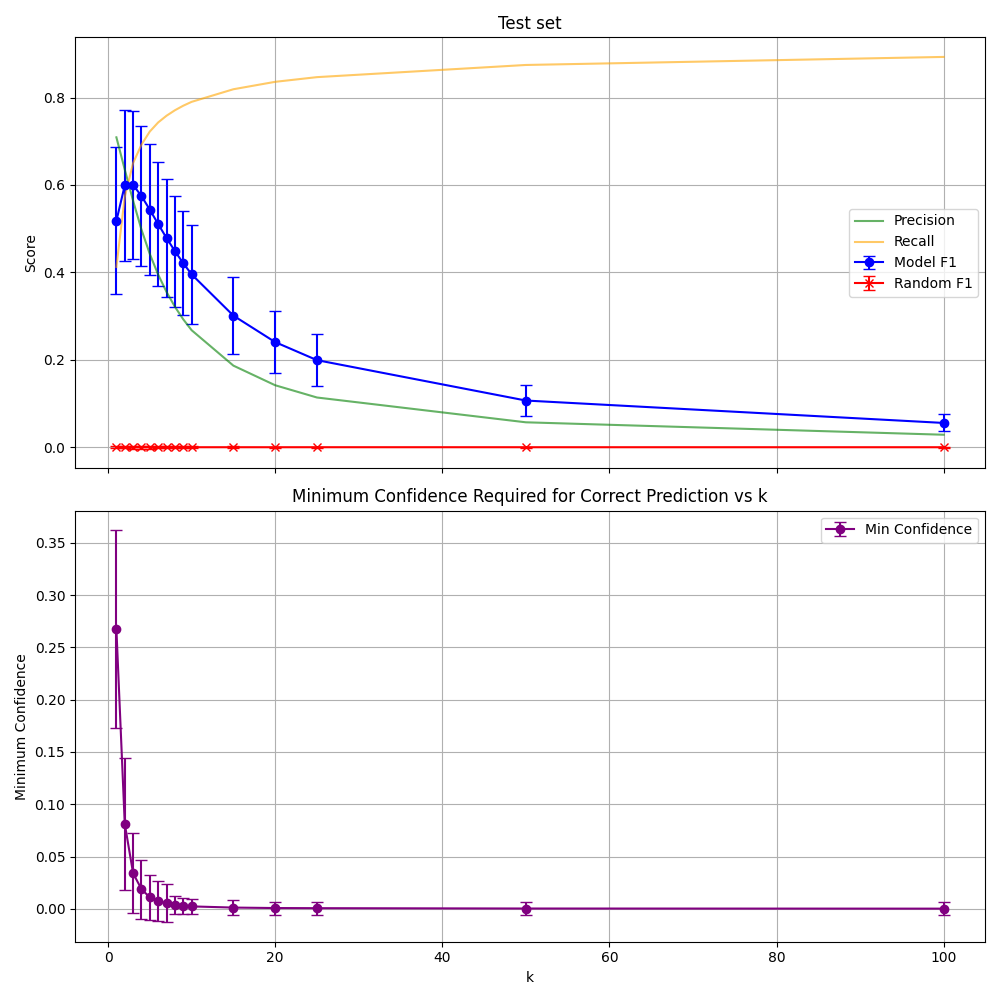}
    \caption{}
  \end{subfigure}
  %\hfill
  \begin{subfigure}[t]{0.34\textwidth}
 %   \centering
    \includegraphics[width=\textwidth]{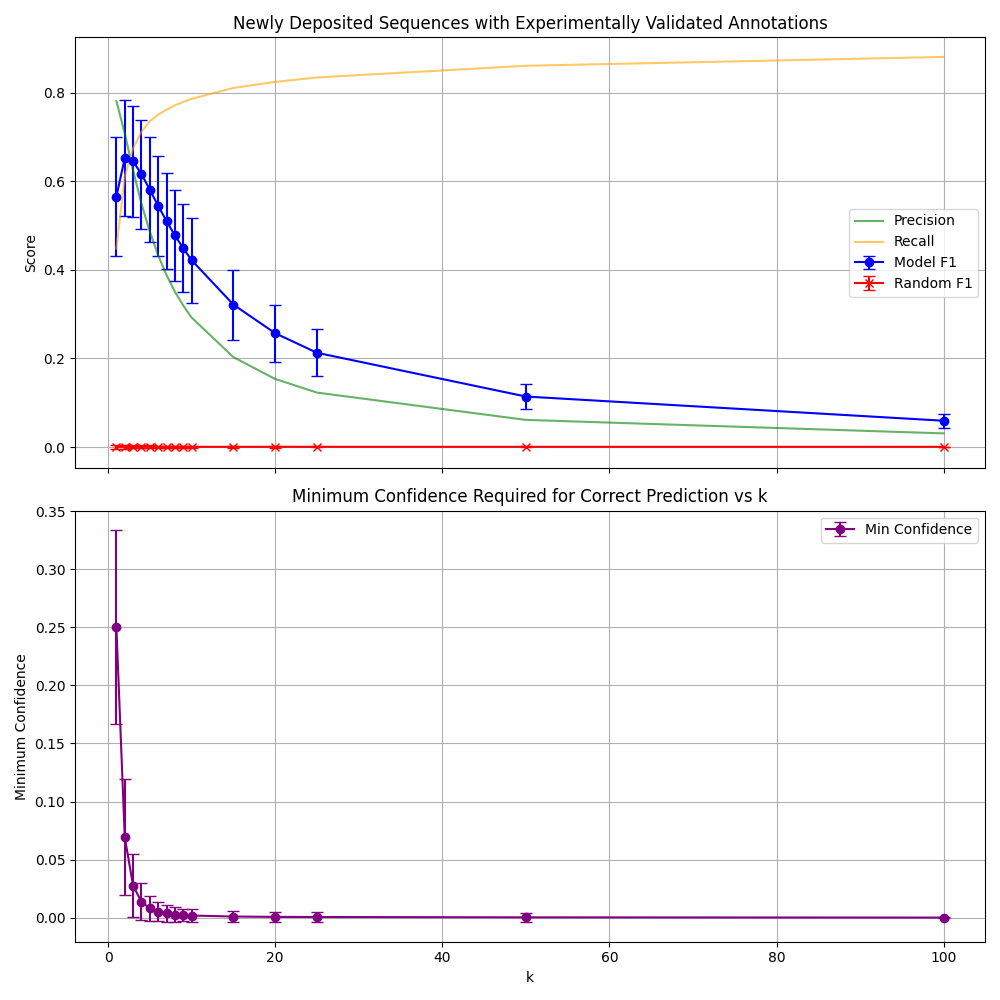}
    \caption{}
  \end{subfigure}
  \caption{Reported precision, recall, and F1 scores for \textit{Translator} while varying $k$ (logit-retrieval) from 1-100, as well as the minimum confidence for correctness. (a) Test set performance. (b) Case study performance.}
  \label{fig:supp_Translator}
\end{wrapfigure}

Translator is a transformer-like network that directly translates protein primary sequence to AV terms. It was used in this study to estimate the protein functions of natural and generated sequences. It outputs any predicted ``aspect'' information, including Enzyme Commission (EC) numbers, Gene Ontologies (GO) for Biological Process (BP), Molecular Function (MF), and Cellular Compartment (CC), in addition to Interpro (IP) domains, Gene3D (3D) domains, UniProt keywords, and cofactors. To support the reliability of the closed-source \textit{Translator} model in our work, we report performance metrics on the Translator test set and a recent case study on its performance. The test set was a non-redundant split of 1,000 sequences and annotations from high-quality UniProt entries that the model was not trained on. Additionally, the case study consisted of 656 UniProt entries deposited after \textit{Translator} was trained that have experimentally verified annotations. \textbf{Supplemental Table \ref{tab:supp_Translator_table}} reports an aspect-by-aspect performance for the test set and case study set using the default settings in the Synthyra API. We also examined the same performance averaged together, varying the internal parameter top-$k$, which allowed \textit{Translator} to be more exploratory with better recall or more precision (\textbf{Supplemental Figure \ref{fig:supp_Translator}}). The default parameter was $k=3$, which had the highest F1 score. We tracked the minimum confidence in prediction such that every prediction above that confidence was correct for each $k$.

\begin{table}[b]
  \centering \small
  \caption{Performance of each \textit{aspect} of \textit{Translator} outputs. (a) Test set metrics. (b) Case study metrics.}
  \begin{minipage}[t]{0.48\textwidth}
    \centering
    \textbf{(a) Test set metrics}
    \begin{tabular}{lcccc}
      \toprule
      \textbf{Aspect} & \textbf{Precision} & \textbf{Recall} & \textbf{F1} & \textbf{Accuracy} \\
      \midrule
      EC       & 0.21 & 0.85 & 0.33 & 0.85 \\
      BP       & 0.16 & 0.75 & 0.27 & 0.75 \\
      CC       & 0.25 & 0.77 & 0.38 & 0.77 \\
      MF       & 0.37 & 0.73 & 0.49 & 0.73 \\
      IP       & 0.78 & 0.76 & 0.77 & 0.76 \\
      3D       & 0.82 & 0.81 & 0.81 & 0.81 \\
      Keywords & 0.41 & 0.86 & 0.56 & 0.86 \\
      Cofactor & 0.38 & 0.81 & 0.52 & 0.81 \\
      \bottomrule
    \end{tabular}
  \end{minipage}%
  \hspace{0.03\textwidth}
  \begin{minipage}[t]{0.48\textwidth}
    \centering
    \textbf{(b) Case study metrics}
    \begin{tabular}{lcccc}
      \toprule
      \textbf{Aspect} & \textbf{Precision} & \textbf{Recall} & \textbf{F1} & \textbf{Accuracy} \\
      \midrule
      EC       & 0.41 & 0.84 & 0.55 & 0.84 \\
      BP       & 0.15 & 0.56 & 0.23 & 0.56 \\
      CC       & 0.26 & 0.77 & 0.39 & 0.77 \\
      MF       & 0.39 & 0.70 & 0.50 & 0.70 \\
      IP       & 0.95 & 0.81 & 0.88 & 0.81 \\
      3D       & 0.98 & 0.89 & 0.93 & 0.89 \\
      Keywords & 0.56 & 0.84 & 0.67 & 0.84 \\
      Cofactor & 0.71 & 0.67 & 0.69 & 0.67 \\
      \bottomrule
    \end{tabular}
  \end{minipage}
  \label{tab:supp_Translator_table}
\end{table}

\subsection{Protify}
\label{sec:supp_Protify}
Protify (\url{https://github.com/Synthyra/Protify}) is an open-source project that offers low-code pLM analysis. We used it extensively to benchmark the representations of DSM models compared to all pLMs evaluated, as well as to train the secondary structure predictor for our analysis of natural versus generated distributions.

Within Protify, linear neural networks were trained with the architecture:
\[
y = \sigma(\mathrm{LN}(\sigma(\sigma(\mathrm{LN}(x)W_1 + b_1)W_2 + b_2))W_3 + b_3)W_4 + b_4,
\]
mapping input $x \in \mathbb{R}^d$ to output $y \in \mathbb{R}^o$ where $o$ was the number of classes in the dataset. $W_1 \in \mathbb{R}^{d\times h}$, $W_2 \in \mathbb{R}^{h\times h}$, $W_3 \in \mathbb{R}^{h\times p}$, $W_4 \in \mathbb{R}^{p\times o}$, $b$ corresponded to the associated bias term in the linear layer, $\sigma$ was ReLU \cite{relu}, $h$ was 8,192, $p$ was calculated with
\[
\left\lfloor \frac{2 \cdot o + 255}{256} \right\rfloor \cdot 256,
\]
and a 20\% random dropout was applied before every MatMul (except the first one) \cite{dropout}. Cross-entropy loss was used for multiclass and binary datasets, and binary-cross-entropy was used for multi-label datasets. 
 Our training scheme followed 1,000 linear warmup steps from zero to a learning rate of $1e^{-4}$, a cosine learning rate scheduler \cite{cosine_lr}, a batch size of 64, with evaluation on the validation set every epoch. Training occurred until a patience of 10 was exceeded for the validation loss. AdamW was used for the optimizer \cite{adamw}. Weighted F1 scores were reported for the test sets in the results.

\subsection{Supervised datasets}
\label{sec:supervised_data}

We used the supervised datasets without modification to the splits or labels except for secondary structure (SS). For SS3 and SS8 we used the Proteinea training set for training \cite{elnaggar_ankh_2023}, CB513 and TS115 for validation \cite{CB513, TS115}, and CASP12, CASP13, and CASP14 for testing. Instead of the common approach to not use intrinsically disordered residue labels, we created a new label for those residues (D) \cite{AV}. As such, there were four and nine classification options per residue for SS3 and SS8, respectively. We sometimes referred to them as \textbf{SS4} and \textbf{SS9}. EC, CC, MF, BP, DL2, DL10, and MB were downloaded from the SaProt repository \cite{su_saprot_2023}. 

\begin{figure}
\centering
\begin{itemize} 
\begin{boxA} %\vspace{-0.5em}
\noindent{\footnotesize
    \item \underline{EC}: 13.1k train, 1.5k valid, 1.6k test - multi-label classification - 585 classes
    \item \underline{CC}: 26k train, 3k valid, 3.4k test - multi-label classification - 320 classes
    \item \underline{MF}: 26k train, 3k valid, 3.4k test - multi-label classification - 489 classes
    \item \underline{BP}: 26k train, 3k valid, 3.4k test - multi-label classification - 1943 classes
    \item \underline{DL2}: 5.5k train, 1.3k valid, 1.7k test - Binary classification
    \item \underline{DL10}: 8,7k train, 2.2k valid, 2.8k test - Multiclass classification - 10 classes
    \item \underline{MB}: 5k train, 662 valid, 665 test - Binary classification
    \item \underline{HPPI}: 26.3k train, 234 valid, 180 test - Binary classification of two input sequences
    \item \underline{SS4}: 10.8k train, 626 valid, 50 test - Token-wise classification - 4 classes 
    \item \underline{SS9}: 10.8k train, 626 valid, 50 test - Token-wise classification - 9 classes 
}
\end{boxA}
\end{itemize}
 \caption{Supervised datasets used to probe model performance. EC, CC, MF, BP, DL2, DL10, MB, and HPPI were from the SaProt repository \cite{su_saprot_2023}. SS4 and SS9 were modified from Proteinea \cite{elnaggar_ankh_2023}.}
  \label{fig:supp_data}
\end{figure}

\subsection{Literature review for known binders of BenchBB}
\label{sec:lit_review}

\begin{sidewaystable}[htbp]
  \centering \small
  \caption{Literature-derived target-binder attributes and corresponding binder sequences. Binding affinity is pKd.}
  \begin{tabularx}{0.98\textheight}{@{} 
      l 
      p{1.8cm} 
      p{3.5cm} 
      p{1.2cm} 
      p{1.1cm} 
      p{2.2cm} 
      >{\RaggedRight\arraybackslash}X 
    @{}}
    \toprule
    \textbf{Target} 
      & \textbf{Target Length} 
      & \textbf{Binding Residues} 
      & \textbf{Binder Length} 
      & \textbf{pK\textsubscript{d}} 
      & \textbf{Binder Source} 
      & \textbf{Binder Sequence} \\
    \midrule

    \textbf{EGFR} 
      & 621 
      & S11, N12, K13, T15, Q16, L17, G18, S356, S440, G441 
      & 241 
      & 8.92 
      & \cite{egfr_comp} 
      & \texttt{\seqsplit{QVQLQQSGPGLVQPSQSLSITCTVSGFSLTNYGVHWVRQSPGKGLEWLGVIWSGGNTDYNTPFTSRLSISRDTSKSQVFFKMNSLQTDDTAIYYCARALTYYDYEFAYWGQGTLVTVSAGGGGSGGGGSGGGGSDILLTQSPVILSVSPGERVSFSCRASQSIGTNIHWYQQRTNGSPKLLIRYASESISGIPSRFSGSGSGTDFTLSINSVDPEDIADYYCQQNNNWPTTFGAGTKLELK}} \\
    \addlinespace

    \textbf{BBF-14} 
      & 122 
      & -- 
      & 147 
      & 7.68 
      & \cite{epfl_bindcraft} 
      & \texttt{\seqsplit{SPIQEEIQKKVRELLEKLIEYLEELKEKAKPPFKEKLEEVIEGLERLKEEVDKVQLNMNLIVFEGLEVDEEGRVWFIVKEMLHATTEEEALENMDKFLESWEKVFKELLEYHFEHNDTSPTFDFFLDFLWWQLYGEPMPKGSHHHHH}} \\
    \addlinespace

    \textbf{BHRF1} 
      & 159 
      & E89, L98, G99, R100  
      & 125 
      & 8.07 
      & \cite{alphaproteo} 
      & \texttt{\seqsplit{MPSAFQIGLALVAAALDRALPEPYRGLALAIAAELSGLPEEELRRLVEAAEKAASADLPFEQQVGLALARIAAAVAGVGLARRAPSLPPEELLAAIREAIEEGGRIAAKALTRSGALEPVLAELP}} \\
    \addlinespace

    \textbf{SpCas9} 
      & 1368 
      & T360 
      & 116 
      & 6.42 
      & \cite{epfl_bindcraft} 
      & \texttt{\seqsplit{SEEEKKEILYFIMEKLFDLDFNFKWPRNSPEEYTKAIEEFKAEVAKIVLETKEKFPEISPEELVELLEEAVYRVHRITHHWASYYVAREVIYELKKLKEKGWKAIEEYTESIISKV}} \\
    \addlinespace

    \textbf{IL7R$\alpha$} 
      & 219 
      & V58, L80, Y139 
      & 64 
      & 10.08 
      & \cite{alphaproteo} 
      & \texttt{\seqsplit{MTKVEEAKELVDKIMEAAKAKDLEKVNKLRTEFFELVNSLSLEEAEEVRKYADKKGEEWYKEQL}} \\
    \addlinespace

    \textbf{MBP} 
      & 370 
      & Y90, P91, F92, Y171, Y176, P315, M321, I329 
      & 126 
      & 7.62 
      & \cite{sybody_mbp}
      & \texttt{\seqsplit{SQVQLVESGGGSVQAGGSLRLSCVASGDIKYISYLGWFRQAPGKEREGVAALYTSTGRTYYADSVKGRFTVSLDNAKNTVYLQMNSLKPEDTALYYCAAAEWGSQSPLTQWFYRYWGQGTQVTVSA}} \\
    \addlinespace

    \textbf{PD-L1} 
      & 221 
      & I54, Y56, E58, N63, Q66, V76, R113, M115, S117, A121, Y123 
      & 120 
      & 10.38 
      & \cite{alphaproteo} 
      & \texttt{\seqsplit{SAEEKILANLEAMKAKALAAKTEEEKLFYAKALLAVAISYAIRGDYELARRAAELAVEVIKSLSKEEQKKVMDFLINIIKNITDPEDREKAIELAIAIAERLDEEVREEALKKIEELKKE}} \\

    \bottomrule
  \end{tabularx}
  \label{tab:supp_benchbb}
\end{sidewaystable}

We conducted an extensive literature review to identify the current highest-affinity binders for each of the BenchBB targets. Binder sequence, binding residues, and affinity data come from the sources reported in \textbf{Supplementary Table \ref{tab:supp_benchbb}}, unless otherwise specified. Sources included Adaptyv Bio's crowdsourced protein design preprint \cite{egfr_comp}, AlphaProteo \cite{alphaproteo}, and BindCraft \cite{epfl_bindcraft}. Although Adaptyv's materials provided Protein Data Bank (PDB) IDs for each target, exact target protein sequences were not explicitly defined. We used sequences for binding prediction using the following rationale:

\begin{itemize}
    \item \textbf{EGFR:} We used the target sequence directly from Adaptyv's Protein Design Competition.
    \item \textbf{SpCas9:} We used the full canonical sequence from UniProt (Q99ZW2).
    \item \textbf{BBF-14:} A de-novo designed protein. We sourced the sequence from PDB (9HAG).
    \item \textbf{BHRF1:} The target sequence was obtained from PDB (2WH6). While the binder sequence with the lowest K\textsubscript{d} came from \cite{alphaproteo}, the binding residues reported in the table came from \cite{bhrf1_binding_sites, bhrf1_more_binding}. The binding residues reported in \cite{alphaproteo} are F65, T74, E77, D82, S85, R93.
    \item \textbf{IL7R$\alpha$:} The sequence was obtained from PDB (3DI3). \cite{alphaproteo} specified that the synthesized protein used in their work was residues 21-239. Upon comparing the PDB entry with the canonical UniProt sequence (P16871), we observed that they were identical from residue 21 to the C-terminus. The FASTA sequence reported in 3DI3 contained a cloning artifact at the N-terminus (``GSHM``), which was removed.
    \item \textbf{MBP:} We used the canonical UniProt sequence (P0AEX9), removing the initial 26 amino acids to remain consistent with the binding residue information reported in \cite{mbp_binding}.
    \item \textbf{PD-L1:} We obtained the sequence from PDB (4Z18). To remain consistent with the binding residue information, retrieved from \cite{pdl1_binding_sites}, we remove the first amino acid.
\end{itemize}

Full details are shown in \textbf{Supplementary Table \ref{tab:supp_benchbb}}.

\subsection{Raw and supporting data}

\begin{figure}[htbp]
  \centering
  \includegraphics[width=.95\textwidth]{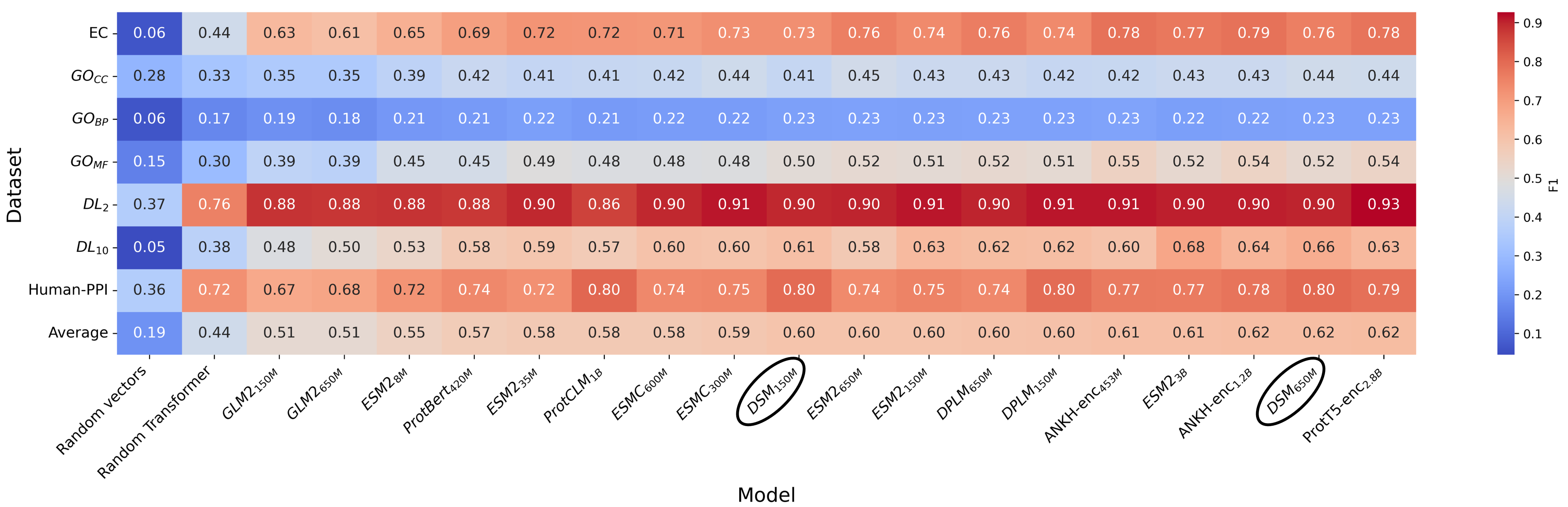}
  \caption{Raw weighted F1 scores (F1$_{max}$ for multilabel problems) of linear probes for pLMs and datasets evaluated for representation quality.}
  \label{fig:supp_rep}
\end{figure}

\begin{figure}[htbp]
  \centering
  \includegraphics[width=.85\textwidth]{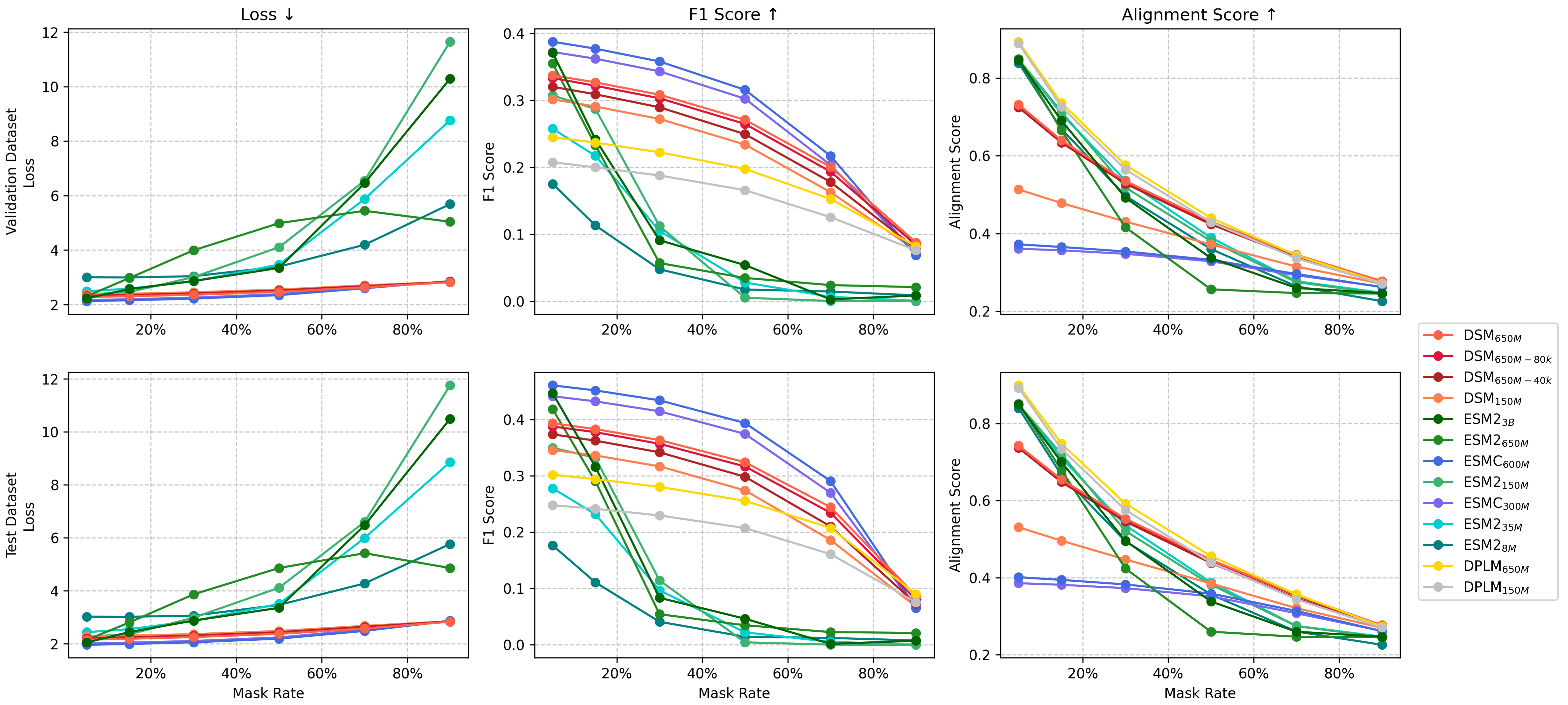}
  \caption{Identical mask rate comparison plot to \textbf{Figure \ref{fig:mask_fill}} except with checkpoints for DSM$_{650}$ and ESMC models added \cite{esm2024cambrian, ESM++}. The 40k and 80k step checkpoints demonstrate the gradual increase in performance of DSM over the training period. The ESMC models show exceedingly high performance across the board, although DSM still produces sequences with better ASc. However, this did not surprise us: while not much is known about ESMC training, it was pretrained on extensive meta-genomic data \cite{esm2024cambrian}, and it is likely that it was trained on our evaluation sets. Additionally, the scaling laws of ESMC sequence reconstruction performance look much closer to the diffusion models than the MLM-based ESM2. This leads us to postulate that ESMC had a dramatically altered training scheme vs. MLM, perhaps directly a diffusion process or some higher or varied mask-rate MLM.}
  \label{fig:supp_mask_fill_esmc}
\end{figure}

\begin{figure}[htbp]
  \centering
  \includegraphics[width=.8\textwidth]{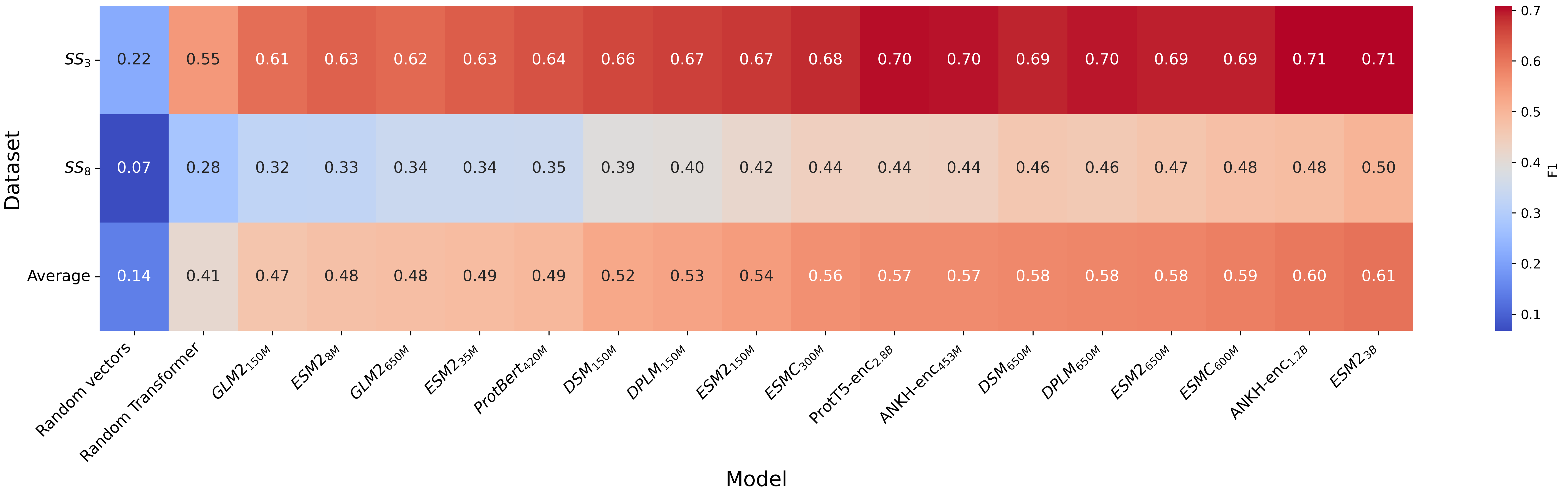}
  \caption{Transformer probe probing of various pLMs on secondary structure datasets, weighted F1 scores reported.}
  \label{fig:supp_ss_screen}
\end{figure}

\begin{figure}[htbp]
  \centering
  \setlength\fboxrule{1.5pt} % Set border thickness
  \begin{subfigure}[t]{0.37\textwidth}
    \centering
    \caption{}
    \vspace{1em}
    \fbox{\includegraphics[width=0.95\textwidth]{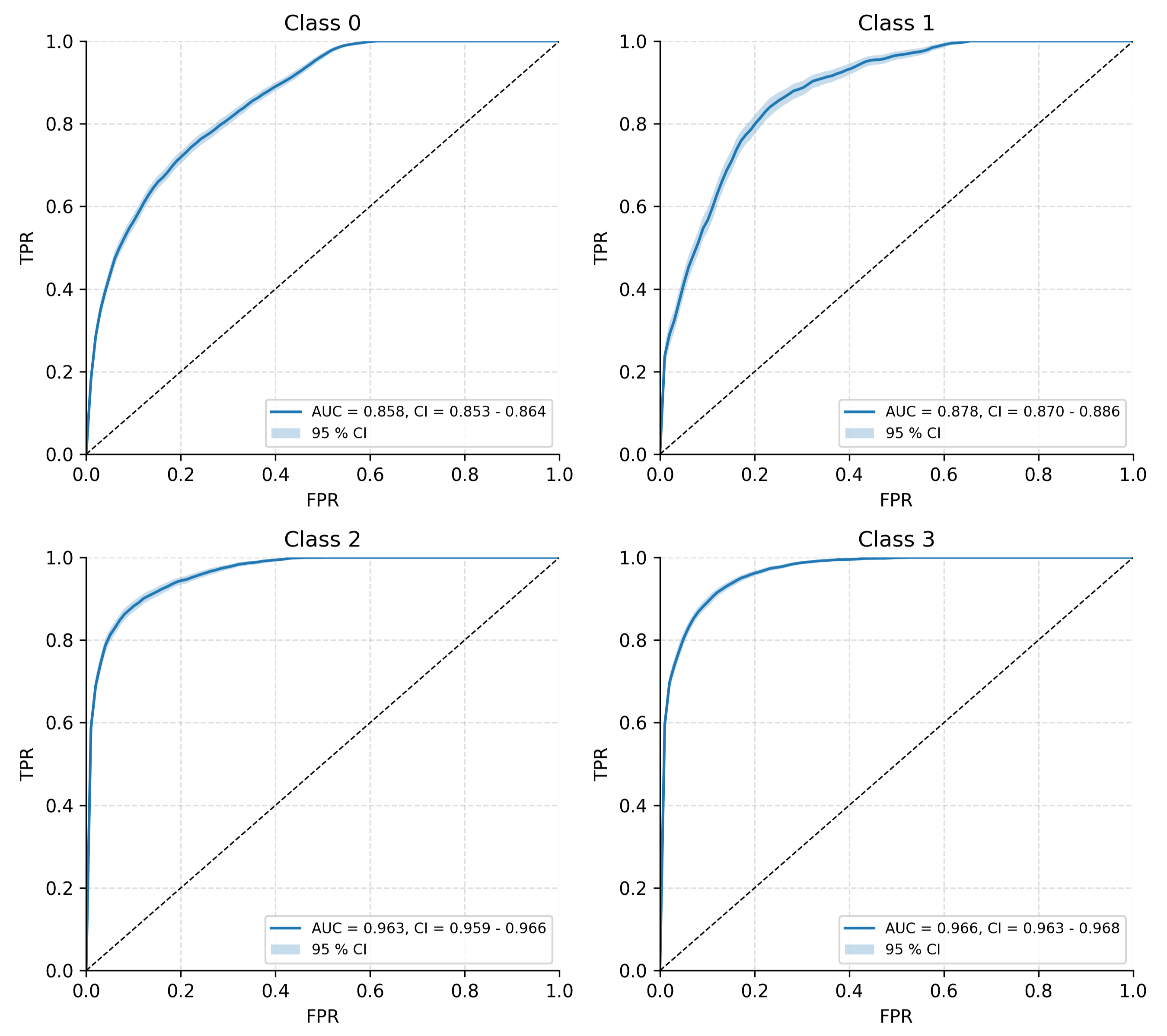}}
  \end{subfigure}
  \hspace{1em}
  \begin{subfigure}[t]{0.55\textwidth}
    \centering
    \caption{}
    \vspace{1em}
    \fbox{\includegraphics[width=0.95\textwidth]{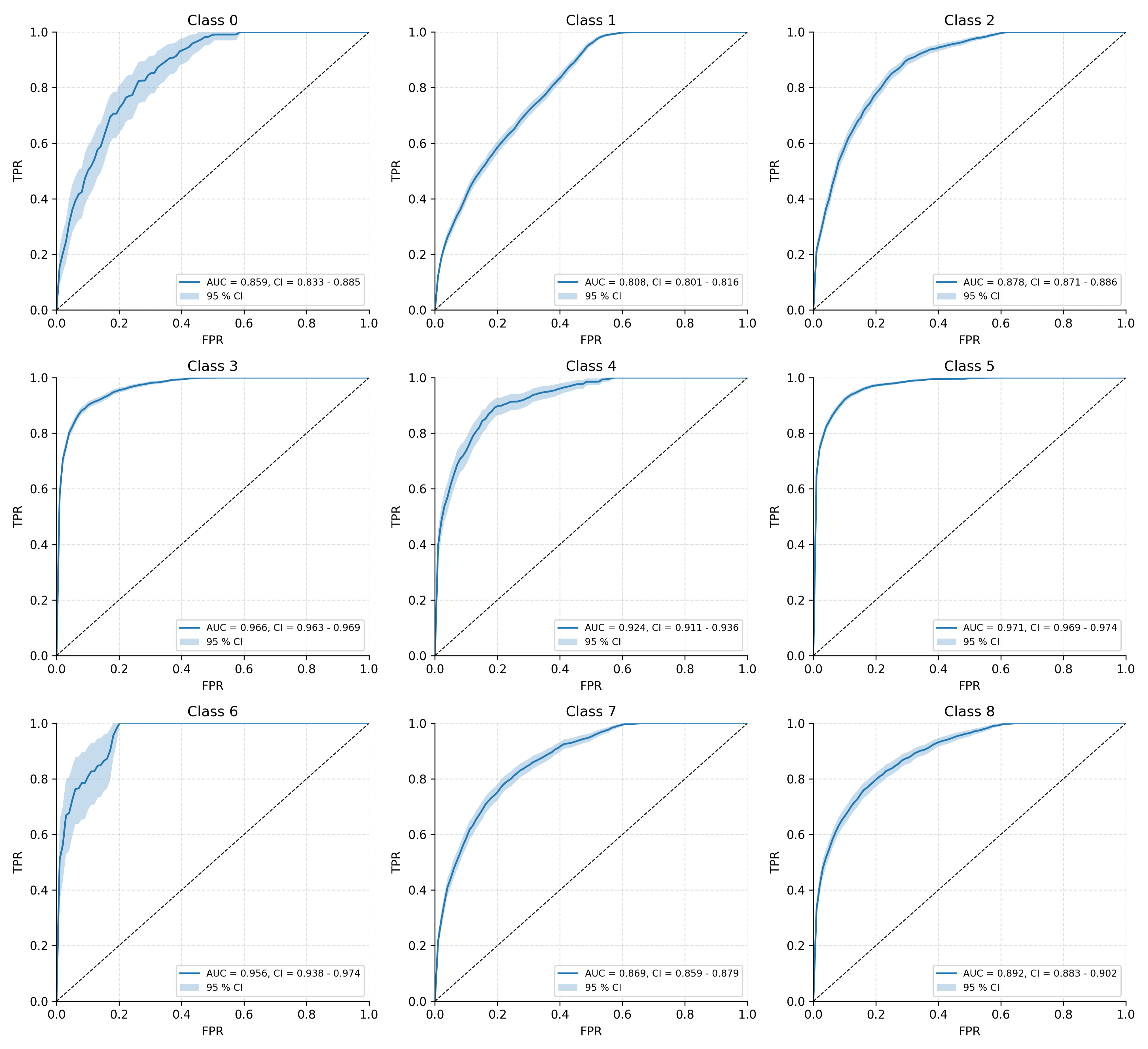}}
  \end{subfigure}
  \caption{
  We selected the ESMC-600 (ESM++ large) \cite{esm2024cambrian, ESM++} as the final model for the base of our production model, balancing performance with throughput using \textbf{Supplemental Figure \ref{fig:supp_ss_screen}}. Shown are AUC plots with statistically sound 95\% confidence intervals around an ROC curve added using DeLong's test in the pAUC package \cite{pauc_1, pauc_2}. For reference, the classes in each model represent the following secondary structures (DSSP conventions with D denoting disordered regions): \textbf{SS4}: 0 = C (\textit{coil/loop}); 1 = D (\textit{disordered}); 2 = E (\textit{beta strand}); 3 = H (\textit{alpha helix}), and \textbf{SS9}: 0 = B (\textit{beta bridge}); 1 = C (\textit{coil/loop}); 2 = D (\textit{disordered}); 3 = E (\textit{beta strand}); 4 = G (\textit{3\textsubscript{10} helix}); 5 = H (\textit{alpha helix}); 6 = I (\textit{pi helix}); 7 = S (\textit{bend}); 8 = T (\textit{turn}).
  \textbf{(a)} Four-class secondary structure (SS4) production model (test set). \textbf{(b)} None-class secondary structure (SS9) production model (test set).}
  \label{fig:supp_ss_prod}
\end{figure}

\begin{table}[htbp]
  \centering \small
  \caption{Reported sequence reconstruction metrics on the PPI test set. DSM$_{ppi}$ has SeqA + SeqB inputs, while the remainder only receives SeqB. Metrics reported for a 15\% mask rate.}
  \label{tab:supp_ppi}
  \begin{tabular}{lcccccc}
    \toprule
    \textbf{Model} & \textbf{Cross entropy}$\downarrow$ & \textbf{MCC}$\uparrow$ & \textbf{F1}$\uparrow$ & \textbf{Recall}$\uparrow$ & \textbf{Precision}$\uparrow$ & \textbf{Accuracy}$\uparrow$ \\
    \midrule
    \rowcolor{green!15} ESM2$_{8}$             & 3.047 & 0.073 & 0.112 & 0.118 & 0.223 & 11.8\% \\
    \rowcolor{green!15} ESM2$_{35}$            & 2.498 & 0.179 & 0.221 & 0.233 & 0.286 & 23.3\% \\
    \rowcolor{green!15} ESM2$_{150}$           & 2.202 & 0.276 & 0.318 & 0.328 & 0.366 & 32.2\% \\
    \rowcolor{green!15} ESM2$_{650}$           & 2.407 & 0.260 & 0.307 & 0.307 & 0.378 & 30.7\% \\
    \rowcolor{green!15} ESM2$_{3B}$            & 2.234 & 0.286 & 0.334 & 0.331 & 0.412 & 33.1\% \\
    \midrule
    \rowcolor{orange!15} DSM$_{150}$           & 2.225 & 0.268 & 0.311 & 0.314 & 0.353 & 31.4\% \\
    \rowcolor{orange!15} DSM$_{650}$           & 2.046 & 0.325 & 0.365 & 0.367 & 0.398 & 36.7\% \\
    \midrule
    \rowcolor{red!15} DSM$_{150-ppi-control}$  & 2.199 & 0.274 & 0.315 & 0.319 & 0.358 & 31.9\% \\
    \rowcolor{red!15} DSM$_{150-ppi}$          & 2.165 & 0.285 & 0.326 & 0.330 & 0.367 & 33.0\% \\
    \rowcolor{red!15} DSM$_{650-ppi}$          & \textbf{1.989} & \textbf{0.342} & \textbf{0.382} & \textbf{0.383} & \textbf{0.417} & \textbf{38.3}\% \\
    \bottomrule
  \end{tabular}
\end{table}

\newpage
\begin{table}[b]
  \centering \small
  \caption{Trends in binder generation of DSM$_{650}$ (Unconditional) and DSM$_{650-ppi}$ (Conditional). Reported is the average predicted binding affinity (ppKd), success rate (percentage higher than best known binder ppKd), and the percentage of designs with a ppKd higher than the best known binder.}
  \begin{tabular}{l c c c c c}
    \toprule
    \textbf{Target} &
    \textbf{Template ppKd error} &
    \multicolumn{2}{c}{\textbf{Conditional}} &
    \multicolumn{2}{c}{\textbf{Unconditional}} \\
    \cmidrule(lr){3-4}\cmidrule(lr){5-6}
    & & \textbf{Success rate (\%)} & \textbf{Avg ppKd} & \textbf{Success rate (\%)} & \textbf{Avg ppKd} \\
    \midrule
    EGFR          & 0.09 & 0.59  & 8.05 & 3.55  & 8.25 \\
    BBF-14        & 1.91 & 4.78  & 8.37 & 7.85  & 8.66 \\
    BHRF1         & 1.52 & 88.20 & 7.05 & 87.45 & 7.12 \\
    IL-7R$\alpha$ & 2.99 & 1.75  & 6.17 & 5.90  & 6.33 \\
    MBP           & 0.24 & 31.70 & 7.19 & 38.40 & 7.27 \\
    Cas9          & 1.11 & 44.63 & 7.52 & 59.35 & 7.54 \\
    PD-L1         & 1.22 & 12.64 & 8.51 & 12.75 & 8.47 \\
    \bottomrule
  \end{tabular}
  \label{tab:supp_generation_trends}
\end{table}

\begin{figure}[htbp]
  \centering
  \includegraphics[width=.8\textwidth]{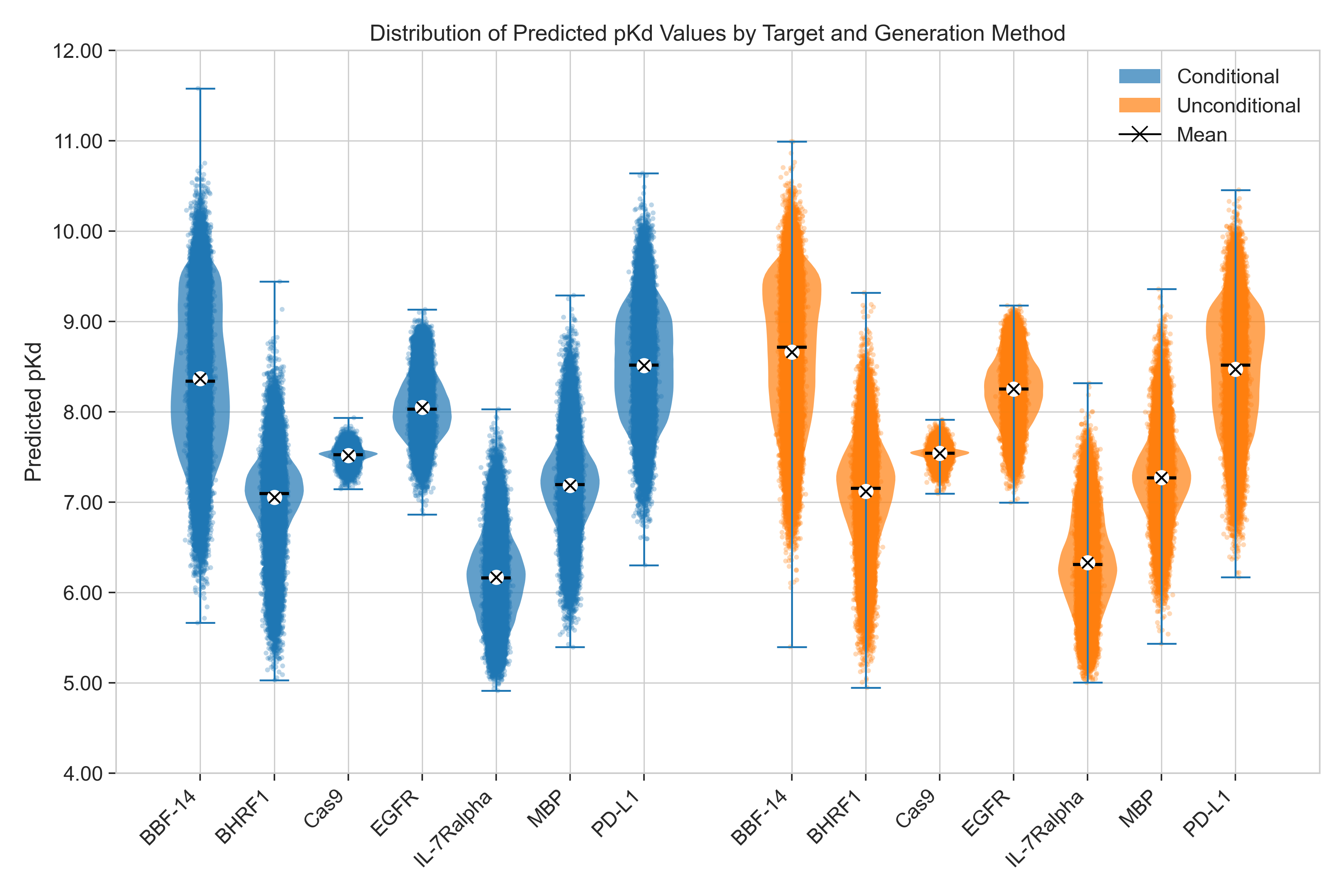}
  \caption{ppKd trends for each protein target for 100,000 designed sequences via Unconditional and Conditional methods.}
  \label{fig:supp_violin}
\end{figure}

\newpage
\begin{figure}[b]
  \centering
  \includegraphics[width=0.8\textwidth]{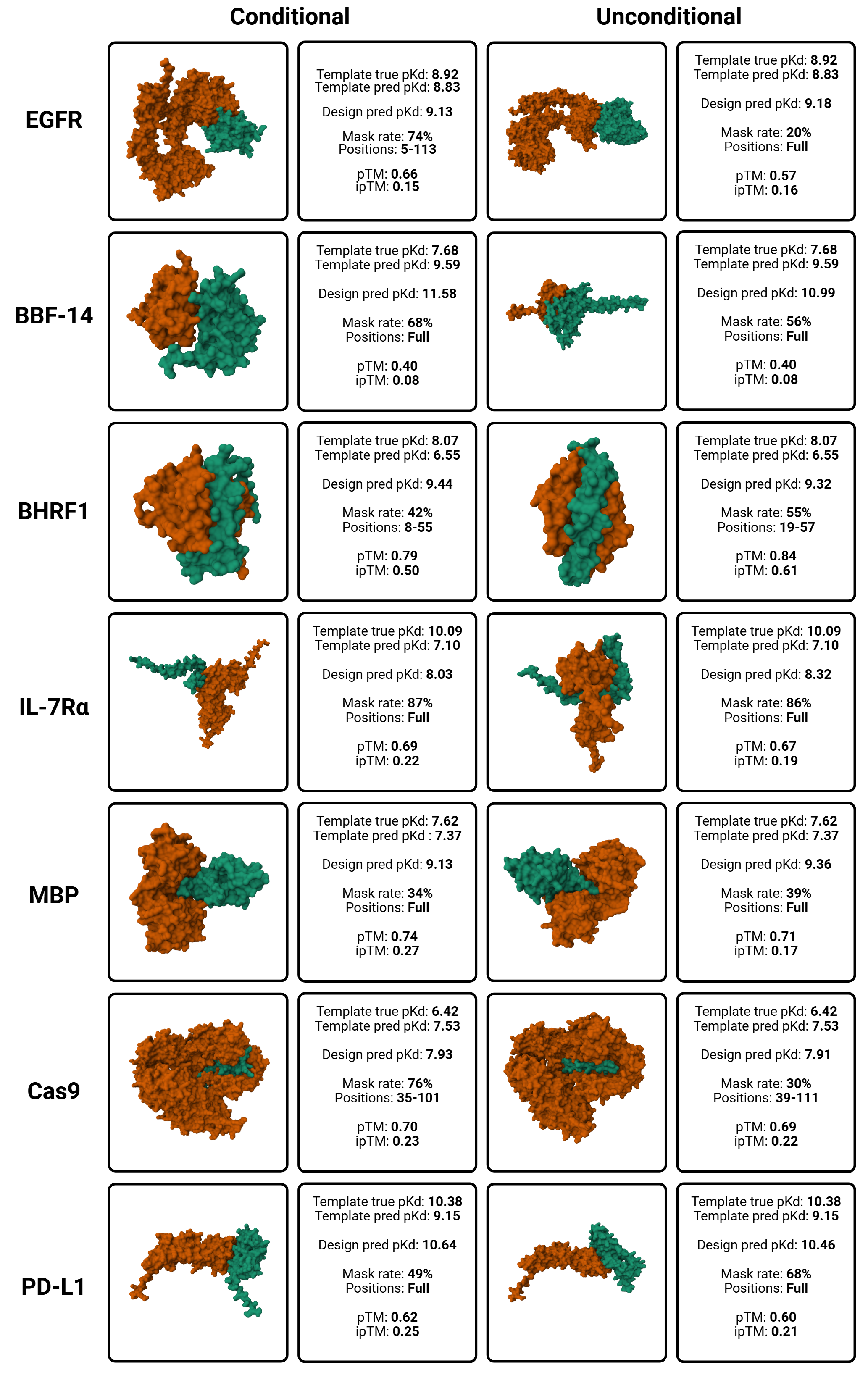}
  \caption{AlphaFold3 folded dimers of BenchBB targets and the best ppKd out of the 100,000 designs via Synteract2 \cite{af3}. Template statistics and AlphaFold3 metrics are reported.}
  \label{fig:supp_best_binders}
\end{figure}

\end{document}